\documentclass[twocolumn]{aastex631}

\usepackage{amsmath}
\usepackage{amsmath,amssymb}
\usepackage{float}
\usepackage[section]{placeins}
\usepackage{multirow}
\usepackage{tablefootnote}
\submitjournal{ApJ}

\graphicspath{{./}{}}
\shorttitle{Burstiness in low stellar-mass H$\alpha$ emitters at  $z\sim2$ and $z\sim 4-6$}
\shortauthors{Navarro-Carrera et al.}

\begin{document}

\title{Burstiness in low stellar-mass H$\alpha$ emitters at  $z\sim2$ and $z\sim 4-6$ from \textit{JWST} medium band photometry in GOODS-S.}

\newcommand{\gsim}{\raisebox{-0.13cm}{~\shortstack{$>$ \\[-0.07cm] $\sim$}}~}
\newcommand{\lsim}{\raisebox{-0.13cm}{~\shortstack{$<$ \\[-0.07cm] $\sim$}}~}

\correspondingauthor{Rafael Navarro-Carrera}
\affiliation{Kapteyn Astronomical Institute, University of Groningen, P.O. Box 800, 9700AV Groningen, The Netherlands}
\email{navarro@astro.rug.nl}
\author[0000-0001-6066-4624]{Rafael Navarro-Carrera}
\affiliation{Kapteyn Astronomical Institute, University of Groningen, P.O. Box 800, 9700AV Groningen, The Netherlands}

\author[0000-0002-5104-8245]{Pierluigi Rinaldi}
\affiliation{Steward Observatory, University of Arizona, 933 North Cherry Avenue, Tucson, AZ 85721, USA}
\affiliation{Kapteyn Astronomical Institute, University of Groningen, P.O. Box 800, 9700AV Groningen, The Netherlands}

\author[0000-0001-8183-1460]{Karina I. Caputi}
\affiliation{Kapteyn Astronomical Institute, University of Groningen, P.O. Box 800, 9700AV Groningen, The Netherlands}
\affiliation{Cosmic Dawn Center (DAWN), Copenhagen, Denmark
}

\author[0000-0001-8386-3546]{Edoardo Iani}
\affiliation{Kapteyn Astronomical Institute, University of Groningen, P.O. Box 800, 9700AV Groningen, The Netherlands}

\author[0000-0002-5588-9156]{Vasily Kokorev}
\affiliation{Department of Astronomy, The University of Texas at Austin, Austin, TX 78712, USA}

\author[0000-0002-1273-2300]{Josephine Kerutt}
\affiliation{Kapteyn Astronomical Institute, University of Groningen, P.O. Box 800, 9700AV Groningen, The Netherlands}

\author{Ryan Cooper}
\affiliation{Kapteyn Astronomical Institute, University of Groningen, P.O. Box 800, 9700AV Groningen, The Netherlands}

\begin{abstract}

We analyze a sample of $\sim 4500$ photometrically-selected H$\alpha$ emitter galaxies at redshifts $z\sim 2$ and $z\sim 4-6$ selected from  James Webb Space
Telescope (JWST) Near-Infrared Camera (NIRCam) medium-band images in the Great Observatories Origins Deep Survey South (GOODS-S). The bulk ($80 \%$) of the galaxies in our sample have stellar masses lower than $\rm 10^{8} \, M_\odot$, with a median stellar mass of $\rm \sim 10^{7.3} \, M_\odot$.
We derive H$\alpha$ rest-frame equivalent widths (EW$\rm_0^{H\alpha}$), line fluxes, and star formation rates using a robust photometric excess technique tested against spectroscopic measurements, being sensitive to EW$\rm_0^{H\alpha} > 75 \, \AA$. Both EW$\rm_0^{H\alpha}$ and $\rm sSFR(H\alpha)$ anti-correlate with stellar mass, and at fixed stellar mass, show a steep increasing trend with redshift $\rm sSFR^{H\alpha} \propto (1+z)^{2.55}$. By comparing the H$\alpha$ and rest-frame UV-derived SFRs, we probe the star formation histories (SFHs) of our galaxies in the past $100$ Myr. The fraction of low-mass galaxies ($\rm \mathcal{M} < 10^{8} \, M_\odot$) with signs of bursty star formation from their SFR(H$\alpha$)/SFR(UV) is $\rm \sim 50 \%$. It quickly drops to $\rm \sim 25 \%$ for $\rm \mathcal{M} > 10^{8} \, M_\odot$. This is consistent with the results from $\rm sSFR(H\alpha)$, showing $80\%$ and $17\%$, respectively. SFR(H$\alpha$)/SFR(UV) is a stricter criterion than those based on the galaxy $\rm sSFR(H\alpha)$, as it only identifies the strongest starbursts, the ones at the initial phases of a bursty star-formation episode.

\end{abstract}
\keywords{High-redshift galaxies (734) -- Galaxy evolution (594) -- Infrared astronomy (786) -- Galaxy photometry (611)}

\newcommand{\xcite}{\textcolor{red}{***}}

%%%%%%%%%%%%%%%%%%%%%%%%%%%%%%%%%%%%%%%%%%%%%%%%%%%%%%
%%%%%%%%%%%%%%%%%% INTRODUCTION %%%%%%%%%%%%%%%%%%%%%%
%%%%%%%%%%%%%%%%%%%%%%%%%%%%%%%%%%%%%%%%%%%%%%%%%%%%%%
\section{INTRODUCTION} \label{sec:intro}

Probing the way in which star formation occurs within galaxies is fundamental to understand galaxy assembly and evolution across cosmic time. The redshift range $z>2$ is particularly important, as it encompasses the period of the first few billion years of cosmic time, in which many galaxies rapidly assembled their stellar mass and reached their chemical maturity. 

Line emitting galaxies are a particularly interesting subgroup of star-forming galaxies. The presence of strong emission lines positively biases their detectability and determination of physical properties, such as their redshift and star formation rate (SFR). This is especially true for H$\alpha$ emitters (HAEs), as H$\alpha$ photons originate in star-forming regions that contain young and massive (O, B) stars, whose lifetime is $\sim 10-20$ Myr. In this way,  the  H$\alpha$ emission line constitutes a direct tracer of the galaxy  `instantaneous' star-formation activity (over the past $\sim 10$ Myr). 

Moreover, the comparison of the SFR derived from  the H$\alpha$ emission with that obtained from other star-formation tracers is a useful tool to constrain the galaxy  star formation history (SFH). Particularly, the galaxy rest-frame UV continuum light is produced mainly by the photospheric emission of stars with masses above $\sim 5 \, M_\odot$ and lifetimes of $\sim 100$ Myr \citep{kennicutt_star_1998}. Thus, the UV-derived SFRs trace the `continuous' star formation (smoothed over the past $\sim 100$ Myr; \citealp{kennicutt_star_1998}). So  the ratio between the SFRs derived from H$\alpha$  and the rest-frame UV indicates how important the instantaneous SF is relative to the smoothed SF in the recent past. Therefore, this ratio can be used as a measure of `burstiness' \citep[e.g., ][]{emami_closer_2019, faisst_recent_2019, atek_star_2022}.

While spectroscopic studies are undoubtedly superior to photometric analyses for characterizing emission lines and galactic physical properties, they are usually limited to the brightest galaxies and/or subject to strong selection effects. So virtually no spectroscopic survey allows for an unbiased line emission study in a stellar-mass selected galaxy sample, especially if it reaches low stellar masses.

To overcome this problem a number of studies have relied on photometric techniques instead to study line emitters.
Studies with ground-based facilities (e.g., \citealp{sobral_large_2013, griffiths_emission_2021}), are usually limited to redshifts where H$\alpha$ emission falls in the observed optical or near-infrared regime, i.e., $z \lesssim 2.5$. Infrared space observatories such as {\it Spitzer} \citep{werner_spitzer_2004} could potentially have traced H$\alpha$ emission up to higher redshifts ($z \sim 8$, \citealp{shim_z_2011, smit_inferred_2016, faisst_coherent_2016, caputi_star_2017}), but due to its lack of sensitivity the H$\alpha$ emission detection was never achieved for individual galaxies, only indirectly via the stacking analysis of relatively massive galaxies \citep{stefanon_high_2022}.

Low-stellar mass galaxies (with $\rm \mathcal{M} \lesssim 10^9 \, M_\odot$) at high redshifts are not only an unexplored realm of the galaxy populations, but also the most numerous ones across all cosmic times \citep{navarro-carrera_constraints_2024}.  It is only now, with the wavelength coverage and sensitivity of the \textit{JWST} instruments, such as the Near-InfraRed Camera (NIRCam) and Near Infrared Spectrograph (NIRSpec), that we can study line emitters amongst low stellar-mass galaxies at high redshifts (e.g., \citealp{rinaldi_midis_2023, rinaldi_midis_2024, caputi_midis_2024, prieto-lyon_production_2023, sun_first_2023}). Particularly the JADES survey (\citealt{eisenstein_jades_2023, eisenstein_overview_2023}) provides us with an ideal dataset for this purpose,  thanks to its combination of depth and area: we can detect galaxies with stellar masses as low as $\rm \mathcal{M} \sim  10^{7} \, M_\odot$ up to $z\sim 7$  over an area of $\rm \sim 60 \, arcmin^2$.

 A number of recent observational and theoretical studies have concluded on the presence of a bursty mode of star formation amongst low stellar-mass galaxies at high redshifts \citep[e.g., ][]{endsley_burstiness_2024, ferrara_eventful_2024}. More systematically, the high incidence of the bursty star-formation mode has been inferred from the study of the location of low stellar-mass galaxies on the SFR-M$^\ast$ plane (\citealp{rinaldi_galaxy_2022,rinaldi_emergence_2024}). In these works, the galaxy SFRs are in general derived from  their rest-frame UV luminosities and starbursts are defined based on an empirical cut on the specific star formation rate ($\rm log_{10} (sSFR/yr^{-1}) > -7.6$;  \citep[][]{caputi_star_2017, caputi_alma_2021}. This cut is physically justified, as it implies a stellar-mass doubling time of $< 4 \times 10^{7} \, \rm yr$, which is consistent with the typical timescales of starburst episodes in the local Universe \citep[][]{leitherer_starburst99_1999}.

In this paper we analyze galaxy burstiness based on the ratio between the SFRs based on the H$\alpha$ luminosity and the rest-UV luminosity. The rationale behind this definition is based on the different SFR timescales that, as we mentioned above, are probed by the two SFR tracers. This criterion has traditionally been applied to galaxy samples up to only intermediate redshifts \citep[e.g.,][]{khostovan_evolution_2024}, with the exception of a few studies at $z>2$ \citep{asada_bursty_2024, endsley_burstiness_2024}. The latter include the analysis of a few hundred low stellar-mass galaxies, down to $\sim 10^8 \, \rm M_\odot$. Here we explore one order of magnitude more galaxies, down to 1 dex lower in stellar mass. 

Our goal is to characterize the dominant mode of star formation and the recent star formation histories of a large sample of low stellar-mass, high-redshift HAEs. For this purpose, we use the JADES (\citealp{eisenstein_overview_2023}) medium-band photometry and the JWST Extragalactic Medium-band Survey (JEMS \citealp{williams_jems_2023}) to select a sample of 4500 H$\alpha$ emitters at redshifts $z\sim 2$ and $z\sim 4-6$ . These datasets allow us to explore stellar masses as low as $\rm 10^{7.5} \, M_\odot$ up to $z \sim 6$, while selecting line-emitters with H$\alpha$ rest-frame equivalent widths (EW$\rm_0^{H\alpha}$) as low as $\rm 75 \, \AA$. 

We adopt throughout a cosmology with $\Omega_{m,0}=0.3$, $\Omega_{\Lambda,0}=0.7$, and $H_0 = $ 70 km s$^{-1}$ Mpc$^{-1}$. Magnitudes are given in the AB system \citep{oke_secondary_1983}. We consider a \cite{chabrier_galactic_2003} initial mass function (IMF) in the stellar mass range of $\mathcal{M}:\rm{0.1\simeq 100} \, {\rm \ M_\odot}$. All the stellar masses and star formation rates taken from the literature have been converted to a \cite{chabrier_galactic_2003} IMF.

\section{DATASET} \label{sec:dataset}
%%%%%%%%%%%%%%%%%%%%%%%%%%%%%%%%%%%%%%%%%%%%%%%%%%%%%%
%%%%%%%%%%%%%%%%%%%% DATASET %%%%%%%%%%%%%%%%%%%%%%%%%
%%%%%%%%%%%%%%%%%%%%%%%%%%%%%%%%%%%%%%%%%%%%%%%%%%%%%%

\begin{deluxetable}{l|ccc}[t]
\tabcolsep=2.25mm
\startdata
\tablecaption{\label{tab:depth} Imaging properties of JADES mosaics. Depths for all bands have been computed following the approach described in Section \ref{subsec:images}, namely $5\sigma$ limiting magnitudes in $0".1$ radius apertures. We divide exposure depths and areas in deep/shallow (in this order, and separated by a slash).}
 & & &\\
 \textbf{Filter} &  \textbf{Depth ($5 \sigma$) } &  Area & Programs\\
 & \textbf{0".1 ap.} & arcmin$^2$ & \\
\hline
F090W & 30.15/29.44 & 30.1/37.5 & JADES\\
F115W & 30.33/29.59 & 35.1/32.5& JADES\\
F150W & 30.30/29.64 & 34.6/32.9 & JADES\\
F182M & 29.75/28.84 & 10.5/52.4 & JEMS+FRESCO\\
F200W & 30.29/29.70 & 36.1/30.5& JADES\\
F210M & 29.74/28.72 & 10.5/50.0 & JEMS+FRESCO\\
F277W & 30.78/30.12 & 35.0/28.4 & JADES\\
F335M & 30.37/30.02 & 17.7/17.7& JADES\\
F356W & 30.95/30.42 & 35.0/27.9& JADES\\
F410M & 30.52/29.80 & 35.5/27.9&  JADES\\
F430M$^\dagger$ & 29.35  & 10.1  &  JEMS\\
F444W & 30.45/29.71 & 28.5/35.0&  JADES+FRESCO\\
F460M$^\dagger$ & 29.07 & 10.1  &  JEMS\\
F480M$^\dagger$ & 29.41 & 10.1  &  JEMS\\
\enddata
\tablenotetext{\dagger}{For these bands, depth is uniform across the entire field, as they come exclusively from JEMS.}
\end{deluxetable}

For this work we use ultra-deep, publicly available JWST/NIRCam data from the JADES survey \citep{eisenstein_jades_2023, eisenstein_overview_2023, rieke_jades_2023, bunker_jades_2023, deugenio_jades_2024}  (\href{https://archive.stsci.edu/hlsp/jades}{DOI: 10.17909/z2gw-mk31}) in GOODS-SOUTH \citep{dickinson_great_2003}. We also make use of ancillary data from the First Reionization Epoch Spectroscopically Complete Observations (FRESCO, PID 1895, \href{https://archive.stsci.edu/hlsp/fresco}{10.17909/gdyc-7g80}, \citealp{oesch_jwst_2023}), and JWST Extragalactic Medium-band Survey (JEMS, PID 1963, \href{https://archive.stsci.edu/hlsp/jems}{10.17909/fsc4-dt61} and \citealp{williams_jems_2023}), alongside all available legacy \textit{HST} images in the field, which we obtained from the Hubble Legacy Fields public images \citep[HLF,][]{whitaker_hubble_2019}.

The JWST/NIRCam bands used in this work are the following: F090W, F115W, F150W, F182M, F200W, F210M, F277W, F335M, F356W, F410M, F430M, F444W, F460M, F480M. The data reduction strategy and procedures are detailed in \citet{rieke_jades_2023} and \citet{eisenstein_jades_2023}. 

In total, we use \textit{9} \textit{HST} and 14 \textit{JWST} (7 medium) bands for our analysis. The combination of both \textit{HST} and \textit{JWST} bands provide deep and almost continuous wavelength coverage from $0.4\mu$m to $4.8\mu$m, which is crucial to assess the physical properties of both host galaxies and emission lines.

\subsection{Imaging Area and Depth} \label{subsec:images}
We determine the depth and area of all images used in our study. We calculate the area of each image by multiplying the total number of non-zero (and non-\texttt{nan}) pixels by the pixel scale of the image.

To assess the depth of each image, we compute the background root mean square (RMS) maps. First, we mask each image using the detection map. Then we randomly place $0".1$ (radius) apertures in the $20\arcsec \times 20 \arcsec$ around each source of the catalog. Finally, we only preserve apertures that fall in areas of empty sky, by ensuring that there is no intersection with any masked region. The standard deviation of the measured fluxes is $1\sigma$ depth in that region of the image. 

The results of this procedure are presented in Table \ref{tab:depth}. Notably, we recovered a $5 \sigma$ limiting magnitude of $\gtrsim 30 \, \rm mag$ for the deep area of JADES bands.

When performing this analysis we found that, the spatial distribution of image depth allows for a division of images in deep/medium areas (see Sect. \ref{appendix} of the appendix). We further validated this approach by checking the distribution of image weights. In most cases, the distribution is bi-modal, with a clear separation between \textit{deep} and \textit{medium} areas.

\section{PHOTOMETRY AND SED FITTING} \label{sec:catalog&sedfit}
%%%%%%%%%%%%%%%%%%%%%%%%%%%%%%%%%%%%%%%%%%%%%%%%%%%%%%%%
%%%%%%%%%%%%%%%%%% catalog&sedfit %%%%%%%%%%%%%%%%%%%%%%
%%%%%%%%%%%%%%%%%%%%%%%%%%%%%%%%%%%%%%%%%%%%%%%%%%%%%%%%
\subsection{Photometric catalog and detection strategy} \label{subsec:catalog}
We use \texttt{SExtractor} \citep{bertin_sextractor_1996} to perform source detection and extract photometry on all images. The detection was performed on an inverse-variance-weighted stacked image composed of F277W, F335M, F356W, F410M, and F444W.

We use a similar detection strategy to the one from \citet{navarro-carrera_constraints_2024} and \citealp{galametz_candels_2013}, but with some modifications in several parameters. In particular, we lowered the detection threshold over the background RMS from $3\sigma$ to $2\sigma$, while increasing the minimum number of pixels to $9$. These allow us to detect faint sources while minimizing the number of spurious detections.

Our catalog includes more than $85\%$ of sources present in the JADES catalog \citep{rieke_jades_2023, hainline_cosmos_2024}. A number of non-JADES-detected (real) sources are present in our catalog, due to the different detection and deblending strategies.

We construct our photometric catalog by measuring $0".25$ (radius) aperture-corrected photometry for sources with \texttt{mag > 27} (\i.e., \texttt{MAG} \_ \texttt{APER}). For brighter sources, we use instead Kron photometry (i.e., \texttt{MAG} \_ \texttt{AUTO}, \citealp{kron_photometry_1980}). Aperture photometry was corrected to total by considering the curve of growth of the PSF (point-spread function) in each filter, using the empirical aperture corrections derived by the JADES team.

We estimate upper limits ($3 \sigma$) by measuring the local background using multiple, circular apertures of $0".25$ (radius), randomly positioned in an $20 \arcsec \times 20 \arcsec$ box around the source. We replaced the flux with the upper limit value and assigned an error of $-1$. We assigned flux and error of \texttt{-99} for sources that are out of a bands coverage.

Finally, we correct all fluxes for Galactic extinction using \texttt{dustmaps} \citet{green_dustmaps_2018}, which is in good agreement with the prescriptions from \citet{schlafly_measuring_2011} \citep{rinaldi_galaxy_2022}. To take into account possible flux calibration uncertainties and because \texttt{SExtractor} is known to underestimate photometric errors (e.g. \citealp{sonnett_testing_2013}), we imposed a minimum error of $0.05$~mag to all the photometry, which also accounts for possible errors in the flux calibration.
\begin{deluxetable}{l|c}[t]
\startdata
\tabcolsep=2.75mm
\tablecaption{\label{tab:sedfit_params} Parameter set employed to retrieve physical properties using \texttt{LePHARE} with \texttt{BC03} stellar population models.}
 & \\
\textbf{Parameter} &  \\
\hline
Templates & \citet{bruzual_stellar_2003}\\ 
SFH & exponential\\ 
$e-$folding time ($\tau$) & 0.01–15 (8 steps) + SSP\\
\tableline
Metallicty (Z/Z$_{\odot}$) & 0.2; 1.00\\
Age (Gyr) & 0.001 – 13.5 (49 steps)\\
\tableline
Extinction laws & \citet{calzetti_dust_2000} +\\
 & \citet{leitherer_global_2002}\\
E(B-V) & 0.0 – 1.5 (16 steps)\\
\tableline
IMF & \citet{chabrier_galactic_2003}\\
Redshift & Fixed to \texttt{EAZY}\\
& \citet{hainline_cosmos_2024}\\
Emission lines & Yes\\
Cosmology (H$\mathrm{_{0}}$, $\mathrm{\Omega_{0}}$, $\mathrm{\Lambda_{0}}$) & 70, 0.3, 0.7
\enddata
\end{deluxetable}

We checked the consistency of our photometry with the official JADES photometric catalog and self-produced catalogs in GOODS-S for our previous study \citep{navarro-carrera_constraints_2024}. No systematic effects were found.

We removed active galactic nuclei (AGN) contaminants from our sample by crossmatching with the catalogs from \citet{lyu_agn_2022, lyu_active_2024} in GOODS-S. This catalogs consolidate a wide array of AGN selection strategies, spanning from X-ray to radio wavelengths, including new JWST Mid-Infrared Instrument (MIRI) selected AGNs.

Finally, we further cleaned the catalog from stars and compact/bright pixel structures present in JADES images. This has been done by identifying sources that fall out of the expected trend of \texttt{MAG\_APER} versus \texttt{FLUX\_RADIUS} (an estimate of the pixel radius containing half of the source's flux), as performed in \citet{barro_candelsshards_2019}. We then visually confirmed the correct removal of sources for a random selection of artifacts. All in all, our final photometric catalog consists of $85.327$ sources.

\subsection{Photometric redshifts and physical properties}

We derived photometric redshifts using \texttt{EazyPY} (\citealp{brammer_eazy_2008}, \citealp{brammer_gbrammereazy-py_2021}, hereinafter \texttt{EAZY}), which relies on spectral energy distribution (SED) fitting. The photometric catalog was used as input for the code. \texttt{EAZY} can estimate photometric redshifts together with their associated error and confidence ($\chi^2$) for each galaxy. Each \texttt{EAZY} best-fit SED is based on a linear combination of all templates in a given redshift grid.

\begin{figure}[t]
\centering
\includegraphics[width=1\columnwidth]{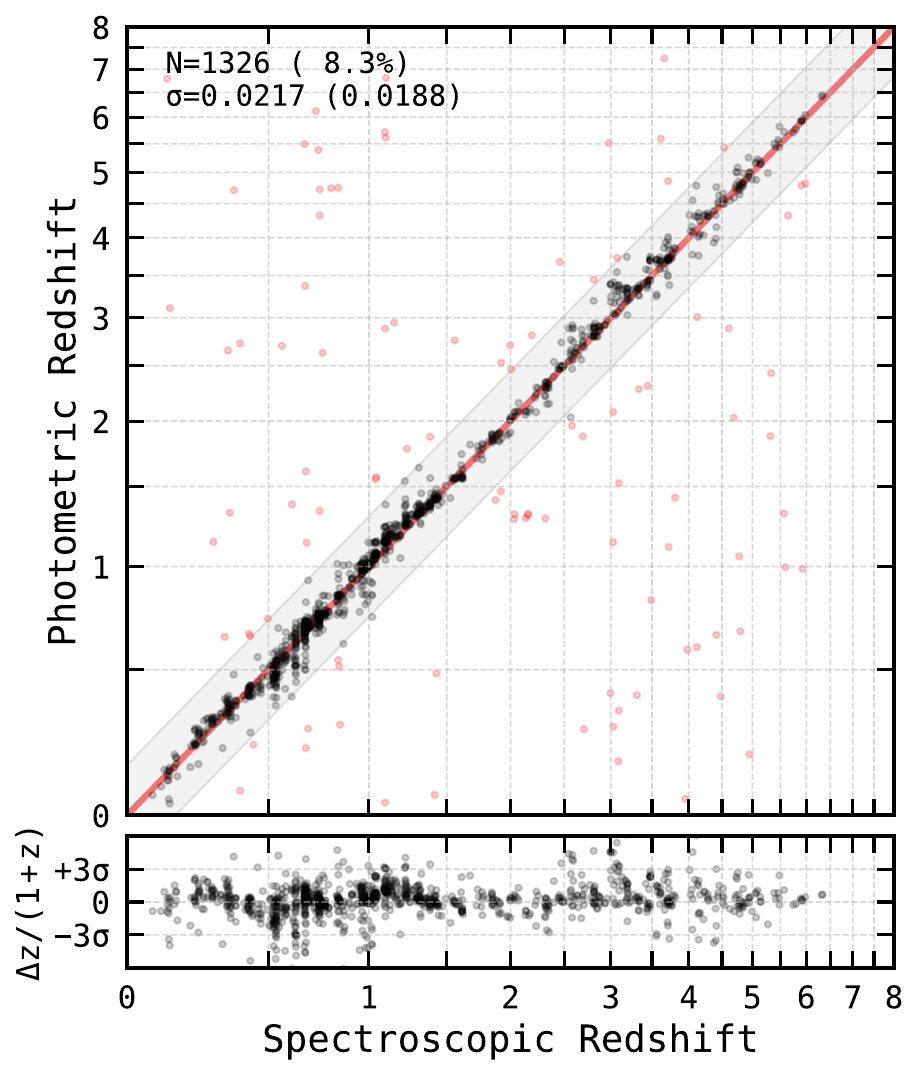}
\caption{Photometric versus spectroscopic redshifts derived using \texttt{EAZY} with the \citealp{hainline_cosmos_2024} templates, based on the fitting of aperture-corrected $0".25$-diameter photometry over the GOODS-S and XDF (excluding galaxies with only FRESCO+\textit{HST} coverage). The compilation of spectroscopic redshifts contains JADES NIRSPec/MSA data \citep{bunker_jades_2023} as well as 3D-HST \citep{brammer_3d-hst_2012} grism results. We include MUSE catalogs in GOODS-S \citep{inami_muse_2017,urrutia_muse-wide_2019,bacon_muse_2023}. Finally, we incorporate the CANDELS GOODS-S catalog \citep{guo_candels_2013}. The sample consists of 1326 pairs, with an outlier fraction of $8.3 \%$ and a standard deviation of $\sigma = 0.0217(0.0188)$ with(without) outliers.}
\label{fig:zspec_zphot}
\end{figure}

We adopted the \texttt{EAZY} template and parameter-set provided by the JADES team (by  \citealp{hainline_cosmos_2024}  and available in \href{https://doi.org/10.5281/zenodo.7996500}{10.5281/zenodo.7996500}). This template set is created starting from the default \texttt{EAZY} templates (\texttt{EAZY-V1.3}), which account for the contribution from nebular emission. The template set includes extra templates that improve the photometric redshift solutions tested against JADES JWST Near Infrared Spectrograph (NIRSec) observations. \citep[see][]{hainline_cosmos_2024}.

Figure \ref{fig:zspec_zphot} shows the \texttt{EAZY}-derived photometric redshifts compared to a collection of $1326$ spectroscopic redshifts from the literature. The outlier fraction (defined as $(z_{phot}-z_{spec}/(1+z_{spec}) > 0.15$) is $8.3\%$ for the sample of $1326$ spectroscopic redshifts, with a normalized median absolute deviation (NMAD) of $\sigma_{NMAD} = 0.02$. This outlier fraction is 2\% higher than the one from JADES photometric redshifts (when using our spectroscopic sample).

We then retrieve the physical properties (stellar mass, best-fit age, metallicity, color-excess and best-fit continuum) for all galaxies in our sample using \texttt{LePHARE} \citep{arnouts_lephare_2011}. We run \texttt{LePHARE} using \texttt{BC03} \citep{bruzual_stellar_2003} templates, after fixing the redshift to the one obtained from \texttt{EAZY}. For the fitting we used \texttt{BC03} models and exponentially declining star formation histories (SFHs) plus a single stellar population (SSP). We allow for solar ($\rm Z_\odot$) and sub-solar ($\rm 0.02 Z_\odot$) metallicities, and ages as young as $10$ Myr. We use the \citet{calzetti_dust_2000}+\citet{leitherer_global_2002} dust reddening law. Table \ref{tab:sedfit_params} presents a detailed description of the adopted parameter set.

\section{Methodology} \label{sec:methods}
%%%%%%%%%%%%%%%%%%%%%%%%%%%%%%%%%%%%%%%%%%%%%%%%%%%%%%%%%
%%%%%%%%%%%%%% selection of ha emitters %%%%%%%%%%%%%%%%%
%%%%%%%%%%%%%%%%%%%%%%%%%%%%%%%%%%%%%%%%%%%%%%%%%%%%%%%%%
In this section we overview the techniques that allowed us to photometrically select our sample of HAEs (\ref{subsec:selection}). Then, we detail the assumptions used to derive star formation rates from the H$\alpha$ (\ref{subsec:ha_sfr}) and rest-frame UV (\ref{subsec:uv_sfr}) luminosities.

\subsection{\label{subsec:selection} Selection of H$\alpha$ emitters using medium band photometry}

As shown by \citet{williams_jems_2023}, the presence of line-emitters in JEMS can be inferred when studying the behavior of NIRCam medium band colors with redshift. A color excess occurs when an emission line falls in one of the bands. As an example, an emission line with $\rm EW_0\sim 100 \, \AA$ falling in F430M would produce a $0.3$ mag photometric excess.

First, we select galaxies with photometric redshifts compatible with finding H$\alpha$ in each of the medium bands in our dataset (see Table \ref{tab:medium_band_ha}). The combination of JADES, JEMS, and FRESCO provides a set of 7 medium bands that span the entire NIRCam wavelength coverage, from $1.8\mu$m to $4.8 \mu$m. The coverage is not continuous, with a gap between F210M and F335M, that prevents us from probing H$\alpha$ between $z\sim2.3$ and $z\sim4$.

From within this subsample of galaxies, we select possible H$\alpha$ emitters (HAEs) using the photometric excess technique \citep[see ][]{rinaldi_midis_2024, caputi_midis_2024}.  We estimate the rest-frame equivalent width of each galaxy using the relation proposed by \citet{marmol-queralto_evolution_2016}:

\begin{equation} \label{eq:ew_0_marmol}
    \rm{EW_0} = \frac{W_{band}}{1+z} (10^{-\Delta \rm{mag}/2.5} - 1).
\end{equation}

Here, $W_{band}$ is the rectangular width \footnote{We defined rectangular width as the length of a rectangle with height equal to the maximum transmission of the filter curve and with the same area as the one covered by the filter transmission curve} of the filter encompassing the flux excess. The excess $\Delta \rm mag = mag_{obs} - mag_{mod} $ is defined as the difference between the observed and model magnitude for the band the emission line falls in (negative when the measured flux density is greater than the expected continuum at that wavelength).

We use the best-fit SED to estimate the expected continuum at the wavelength of each medium band. These fits were performed excluding the band affected by possible emission lines.

We cross-matched our sample with the JADES NIRSpec/MSA catalog \citep{bunker_jades_2023, deugenio_jades_2024} as quality assessments for our HAE selection. The common sample (of 214 galaxies) has an average H$\alpha$ flux of $4.2\times 10^{-18}$ erg/s/cm$^2$. We found a good agreement between photometric and spectroscopic redshifts (outlier fraction of $\sim 5 \%$), significantly smaller than the parent sample.

Our photometrically derived H$\alpha$ fluxes are not statistically significantly different from the spectroscopic ones (from a t-test). The average deviation between our measurements and the spectroscopic ones is $3\times 10^{-19}$ erg/s/cm$^2$.

We imposed the following criteria to select HAEs within our sample:
\begin{enumerate}
    \item mag\_i $\neq -99$ and err\_mag\_i $\neq -1$ for \texttt{band} and all NIRCam wide filters\}
    \item $\Delta \rm mag < min(-0.1,2\times mag\_err\_\texttt{band})$
    \item $\rm SNR$\_\texttt{band}$>3$
    \item $\rm  |\Delta \rm{mag}$\_i $| < 2 \times \rm mag\_err$\_i for i in \{F115W, F150W\} if \texttt{band} is in \{F182M,F210M\} or \{F356W, F444W\} for the rest
\end{enumerate}
Here, \texttt{band} is the filter in which the H$\alpha$ line falls. 

\begin{deluxetable}{l|ccccc}[t]
\tabcolsep=2.25mm
\startdata
\tablecaption{\label{tab:medium_band_ha}  H$\alpha$ $\rm EW_0$ values detected from photometric excess in different NIRCam medium bands .}
 & & &\\
\textbf{Filter} &  $z_{\rm min}$ &  $z_{\rm max}$ & $\rm EW_0^{min}$ ($\rm \AA$) & $\rm \overline{EW}_0$ ($\rm \AA$)\\
\hline
F182M & 1.6 & 2.0 & 78 & 683\\
F210M & 2.0 & 2.3 & 61 & 694\\
F335M & 3.9 & 4.4 & 64 & 574\\
F410M & 4.9 & 5.4 & 67 & 703\\
F430M & 5.4 & 5.7 & 34 & 840\\
F460M & 5.8 & 6.2 & 32 & 970\\
F480M & 6.2 & 6.5 & 41 & 1307\\
\enddata
\end{deluxetable}

By applying this selection, we imposed a minimum photometric offset ($\rm \Delta mag$) of $0.1$ mag (absolute value). This translates directly into a minimum EW$_0$(H$\alpha$) for each medium band, as shown in Table \ref{tab:medium_band_ha}. On average, this value is $\rm EW_0(H\alpha) \sim 50 \, \AA$. The selection also translates into a minimum specific star formation rate (sSFR). Our sample shows a minimum sSFR of $\rm \sim 10^{-8.9} \, yr^{-1}$ (for the galaxies with the highest optical mass-to-light ratio of the sample).

We select around $\sim 3500$ HAEs within our sample, with a median stellar mass of $\rm \sim 10^{7} \, M_\odot$. This corresponds to $\sim 9 \%$ of the sample between the redshifts shown in Table \ref{tab:medium_band_ha}.

\begin{figure}[t]
\centering
\includegraphics[width=1.02\columnwidth]{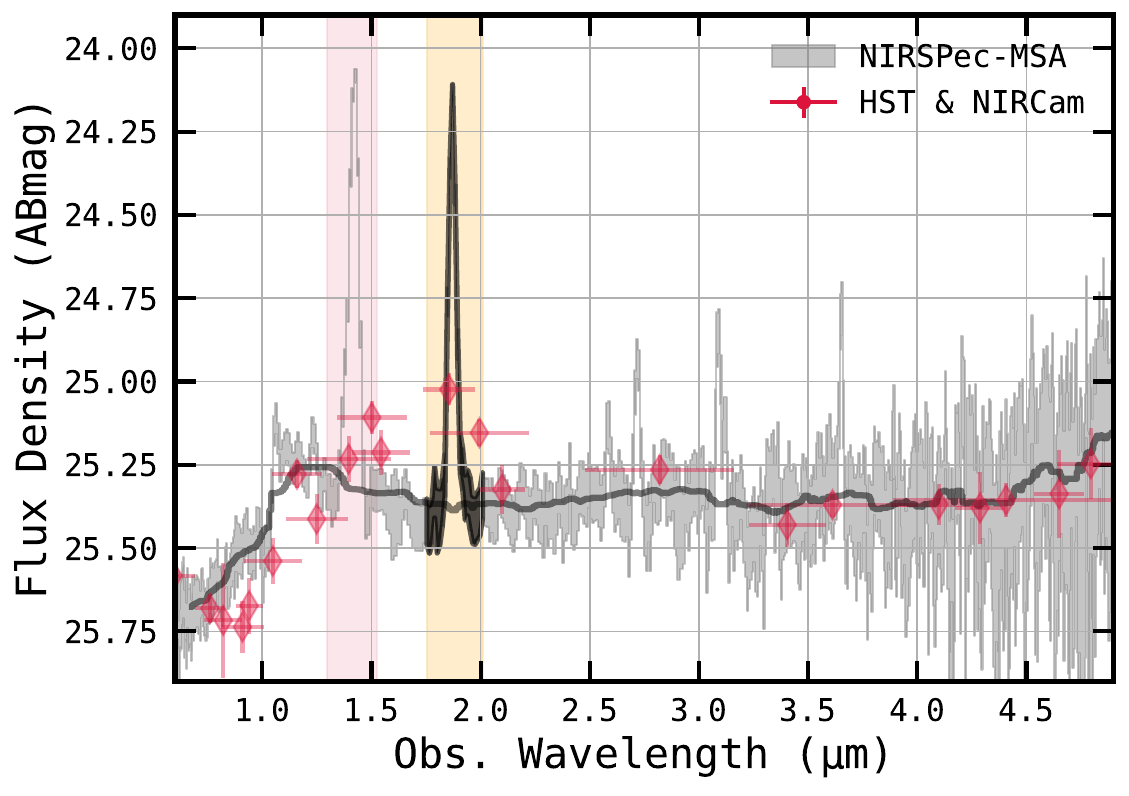}
\caption{\label{Fig:msa_vs_photom} Example of an HAE at $z=1.92$. Photometric observations are shown as red diamonds. The effect of H$\alpha$ and OIII+H$\beta$ emission lines (highlighted with orange and red shades, respectively) can be seen in the photometry, demonstrating the efficiency of the photometric excess methodology. We show the NIRSpec-MSA observations for this source as a grey line, which confirm the presence of strong emission lines as traced from the photometry.}
\label{fig:msa}
\end{figure}

Fig. \ref{Fig:msa_vs_photom} shows an example of an HAE present in both spectroscopic and photometric samples, with $\rm z_{spec} \sim 1.847$. The photometric excess traces both H$\alpha$ (highlighted within an orange area) and (H$\beta$+[OIII]) emission lines (red area), producing a flux density excess above the continuum in the photometric data points of the affected filters.

We emphasize that the photometric-excess technique has been widely used in literature (e.g. \citealp{shim_z_2011, sobral_large_2013, faisst_coherent_2016, marmol-queralto_evolution_2016, rasappu_mean_2016, caputi_star_2017, stefanon_high_2022}) using \textit{SPITZER}, \textit{HST} and ground-based facilities, and more recently \textit{JWST} \citep{caputi_midis_2024, prieto-lyon_production_2023, simmonds_low-mass_2023, van_mierlo_high_2023, rinaldi_midis_2024}.

\subsection{\label{subsec:ha_sfr} Star formation rate from the H$\alpha$ equivalent width}

We can recover the (H$\alpha$+[NII]+[SII]) flux from the photometric excess by using the following relation:
\begin{equation}
    \rm F_{H\alpha+[NII]+[SII]} = F_{cont} \times (10^{-\Delta \rm mag/2.5}-1),
\end{equation}
where $\rm F_{H\alpha+[NII]+[SII]}$ is the flux of the H$\alpha$+[NII]+[SII] complex, and $\rm F_{cont}$ is the continuum flux, as modeled from the best-fit SED. In this equation, $\Delta \rm mag$ has the same meaning as in Eq. \ref{eq:ew_0_marmol}.

We estimated the contribution of H$\alpha$ to the complex  H$\alpha$+[NII]+[SII] by applying the calibration derived by \citet{anders_spectral_2003}. Namely: $\rm F_{H\alpha} = 0.63 \times F_{H\alpha + [NII] + [SII]}$ for solar metallicity ($\rm Z_\odot$), and $\rm F_{H\alpha} = 0.82 \times F_{H\alpha + [NII] + [SII]}$ for sub-solar metallicity ($\rm 0.2 \times Z_\odot$).

We use intrinsic H$\alpha$ luminosities (dust corrected using the Calzetti attenuation law; \citealp{calzetti_dust_2000}) to estimate the galaxy star formation rate \citep[SFR; ][]{kennicutt_star_1998}. We then scale these values to the IMF used in our study \citep{chabrier_galactic_2003},  following the prescription of \citet{madau_cosmic_2014}. In summary, the calibration we used is:
\begin{multline}
    \rm SFR_{H\alpha} \, [M_\odot/ yr] = 7.9 \times 10^{-42}\times \\ \rm \times  0.68  \times L_{H\alpha} [erg \, s^{-1}]
\end{multline}

\subsection{\label{subsec:uv_sfr} Star formation rates from the rest-frame UV luminosities}

We measure the UV luminosity density at 1500 $\AA$ ($\rm L_\nu(1500 \AA)$) and UV absolute magnitude independently from the best-fit SED, by selecting the filter (or filters) with the closest coverage to rest-frame $1500 \rm{\AA}$. We account for dust extinction by using the E(B-V) value given by the best-fit SEDs, then correcting the observed values to intrinsic by using \citet{calzetti_dust_2000} attenuation-law. All values of SFR(UV) and SFR(H$\alpha$) that we discuss in this paper are dust-corrected.

We estimate the UV-SFR by converting the UV luminosity density between $1500-2800 \rm{\AA}$, following the prescription of \citet{kennicutt_star_1998}. This calibration assumes a constant SFR during at least $100$~Myr. The relation we use, including the extra factor to convert from a Salpeter IMF \citep{salpeter_luminosity_1955} to a Chabrier \citep{madau_cosmic_2014} is:
\begin{multline} \label{eq:kennicut}
    \rm SFR_{UV} \, [M_\odot/ yr] =  1.4 \times 10^{-28} \times \\ \rm \times 0.63 \times \, L_\nu(UV) \, [erg \, s^{-1} \, Hz^{-1}]
\end{multline}

As a consequence of the assumptions behind the Kennicutt calibration, the values of the $\rm SFR$ for galaxies that do not have a constant SFH for the last hundred million years differ from the ones that can be recovered by using Eq. \ref{eq:kennicut}. For these galaxies, Eq. \ref{eq:kennicut} can under-predict SFRs by about a factor of two in the case of galaxies with bursty star formation histories (e.g., with a recent burst in the last $10$ Myr, \citealp{kennicutt_star_1998}).

%%%%%%%%%%%%%%%%%%%%%%%%%%%%%%%%%%%%%%%%%%%%%%%%%%%%%%%%%%%%%%%%%%%%%%
%%%%%%%%%%%%%%ew of HAES%%%%%%%%%%%%%%%%%
%%%%%%%%%%%%%%%%%%%%%%%%%%%%%%%%%%%%%%%%%%%%%%%%%%%%%%%%%%%%%%%%%%%%%%
 \section{H$\alpha$ and its relation to the host galaxy properties \label{sec:results_ew}} 

\begin{figure}[t]
\centering
\includegraphics[width=0.95\columnwidth]{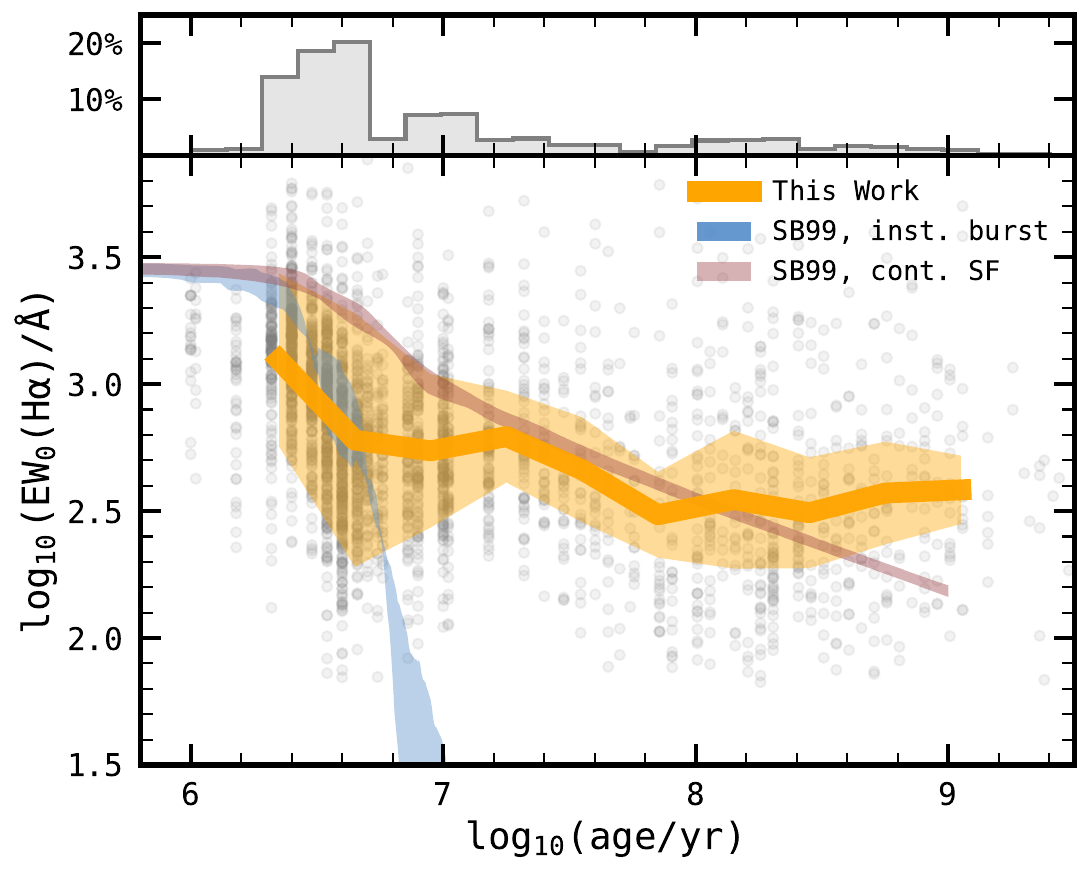}
\caption{\label{Fig:ew_vs_age} Relation between EW$_0$(H$\alpha$) and best-fit age. Grey circles represent the HAEs in our sample. The median trend and $1\sigma$ deviation are included as an orange line and shaded area, respectively. \texttt{Starburst99} tracks for an instantaneous burst and constant star formation are overplotted as blue and pink areas, respectively. We plot them as filled areas, corresponding to range of values between $\rm Z_\odot$ and $\rm 0.2\times Z_\odot$.}
\label{fig:ew_age}
\end{figure}

\begin{figure*}[t]
\centering
\includegraphics[width=0.95\textwidth]{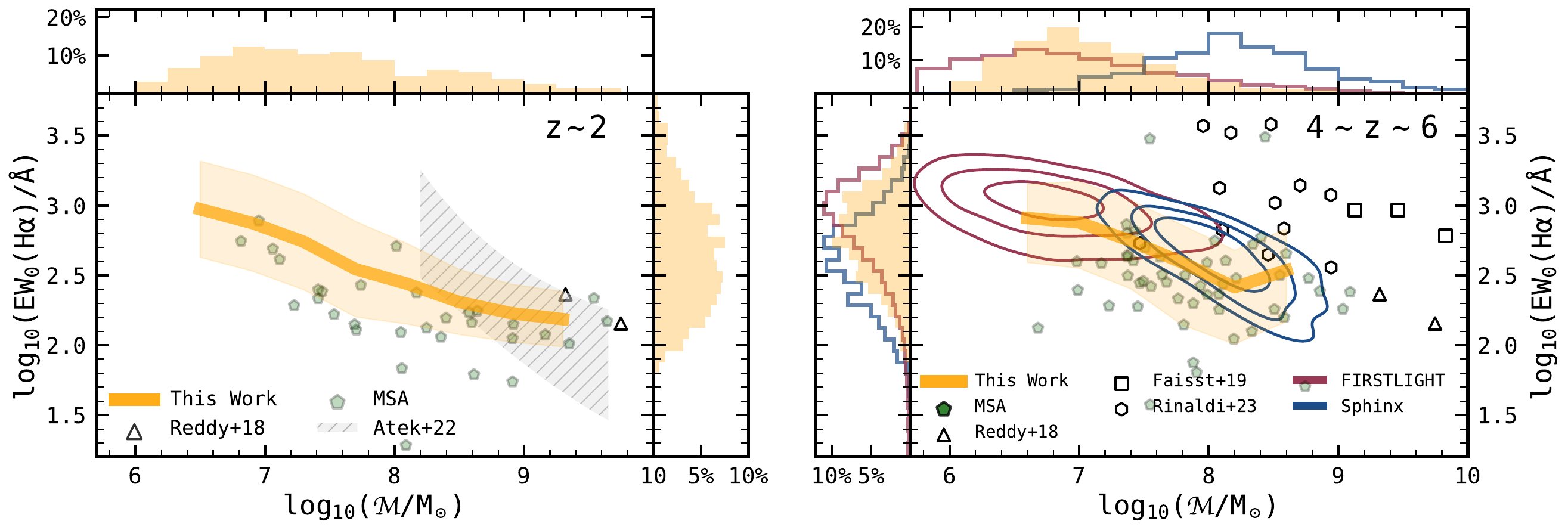}
\caption{\label{fig:ew_vs_m} Relation between EW$_0$(H$\alpha$) and stellar mass. We show our median trend as an orange line, and with a $1\sigma$ shaded area. Spectroscopically selected galaxies from MSA are shown as green pentagons. The \textbf{left panel} displays the $z \sim 2$ sample, also showing the results from \citet{atek_star_2022} as a grey hatched area, and \citet{reddy_mosdef_2018} as triangles. The \textbf{right panel} shows our $z \simeq 4 - 6$ sample, together with the results from \citet{reddy_mosdef_2018}, \citet{faisst_recent_2019} and \citet{rinaldi_midis_2024} (at higher redhisft, $\rm z \sim $) as triangles, squares and hexagons, respectively. SPHINX and FIRSTLIGHT theoretical predictions are shown as blue and red contours.}
\end{figure*}

In this section, we present the relation between EW$_0$(H$\alpha$), age and stellar mass in \ref{subsec:ew_versus_age} and \ref{subsec:ew_versus_mass}; and the evolution of EW$_0$(H$\alpha$) with cosmic time in \ref{subsec:ew_versus_time}. For this analysis, we focus the study on two redshift bins, namely, $z\sim 2$ and $z \sim 4-6.5$ (see Table \ref{tab:medium_band_ha}).

\subsection{\label{subsec:ew_versus_age} The evolution of EW$_0$(H$\alpha$) with age}

It is not surprising that prominent HAEs have relatively young ages, as could be inferred from the presence of a recent star formation burst as massive, short-lived stars are responsible for the ionization of Hydrogen in most of our sample.

In the case of galaxies with a recent burst of star formation, the best-fit age prefers young age solutions. To ensure that these ages are physical, and not merely produced by degeneracies with other parameters (e.g., dust attenuation), we repeat the fit not allowing for ages younger than $300$ Myr. No secondary solution corresponding to a different parameter value combination was chosen for most galaxies.

In Fig. \ref{fig:ew_age} we show the relation between EW$_0$(H$\alpha$) and best-fit age. We include the evolution of EW$_0$(H$\alpha$) for SB99 \citep{leitherer_starburst99_1999} stellar tracks with a single burst and continuous star formation between $\rm Z = 0.2 \times Z_\odot$ and $\rm Z = Z_\odot$.

As found by \citet{caputi_midis_2024} for (H$\beta$+[OIII]) emitters, most galaxies in our HAE sample have young ages ($66\%$ are younger than $100$ Myr). For galaxies younger than $100$ Myr, we find a broad variation of EW$_0$(H$\alpha$), as is expected from a single burst of star formation according to the evolutionary models (blue area). 

We also encounter older galaxies, with a lower average and less scattered EW$_0$(H$\alpha$). This is compatible with a more extended star formation history, or a sustained star formation, as the corresponding stellar track (red area) suggests. 

\subsection{\label{subsec:ew_versus_mass} The evolution of EW$_0$(H$\alpha$) with stellar mass}

Fig. \ref{fig:ew_vs_m} shows the relation between rest-frame EW$_0$(H$\alpha$) and stellar mass. EW$_0$(H$\alpha$) and stellar mass are anti-correlated in our sample, as seen by the running median trend.

We include the JADES-MSA spectroscopic sample of galaxies with robustly available H$\alpha$ flux measurements, obtained from the JADES catalog \citep[second data release,][]{bunker_jades_2023}. This sample broadly follows the median trend of the photometrically-selected HAEs. However, no statistical conclusion has been derived from the spectroscopic galaxy sample, as completeness is not guaranteed down to the lower stellar masses and the selection function is not known.

The range of stellar masses that the HAEs sample spans is $\rm \sim 10^{6} \, M_\odot$ to $\rm 10^{9} \, M_\odot$ (see histograms on the upper part of Fig. \ref{fig:ew_vs_m}). This is at least 1 dex lower than the stellar masses probed in previous works that study HAEs at similar redshifts: the bulk of their galaxy sample has stellar masses in the range of $\rm 10^{8} \, M_\odot$ to $\rm 10^{10} \, M_\odot$ (e.g., \citealp{caputi_star_2017, atek_star_2022, asada_bursty_2024, endsley_burstiness_2024}). 

By using the method presented by \citet{pozzetti_zcosmos_2010}, we have determined our sample to be 75\% complete in stellar-mass down to $\rm \sim 10^{7.25} \, M_\odot$ at $z\sim 6$.

Fig. \ref{fig:ew_vs_m} shows that the median value of EW$_0$(H$\alpha$) for galaxies with a stellar mass of $10^7 \, M_\odot$ at $z\sim2$ is close  to $1000 \, \rm  \AA$, and decreases to $\sim 100 \, \rm \AA$ for galaxies of  $10^9 \, \rm M_\odot$. The situation is similar at $4 \lesssim z \lesssim 6$, altough with a  less steep anti-correlation.

The anti-correlation between stellar mass and EW$_0$(H$\alpha$) has been reported previously in numerous studies of HAEs at different redshifts, using both photometric and spectroscopic samples (\citealp{lee_comparison_2009, fumagalli_h_2012, stark_keck_2013, marmol-queralto_evolution_2016, smit_inferred_2016, reddy_mosdef_2018, atek_star_2022, rinaldi_midis_2024, caputi_midis_2024, llerena_physical_2024}). \citet{sobral_large_2013} determined that the anti-correlation between EW$_0$(H$\alpha$) and stellar mass could be expressed as $\rm EW_0 \propto \mathcal{M}^{-0.25}$, which is in agreement with the results for our sample. \citet{khostovan_correlations_2021} investigated the origin of this correlation. After removing all observational bias, the anti-correlation is still significant, although with a less steep slope. According to our results, this relation still holds for stellar masses $\rm 10^{6} \, M_\odot$ to $\rm 10^{9} \, M_\odot$.

From Fig. \ref{fig:ew_vs_m}, the median trend starts to flatten out around an EW$_0$(H$\alpha$) of $\sim 3000 \, \rm \AA$. The same situation arises for the evolution of EW$_0$(H$\alpha$) with age as predicted by \texttt{SB99}, where EW$_0$(H$\alpha$) also saturates to a maximum value of  $\sim 3500 \, \rm \AA$ for all young ages.

\begin{figure*}[t]
\centering
\includegraphics[width=0.8\textwidth]{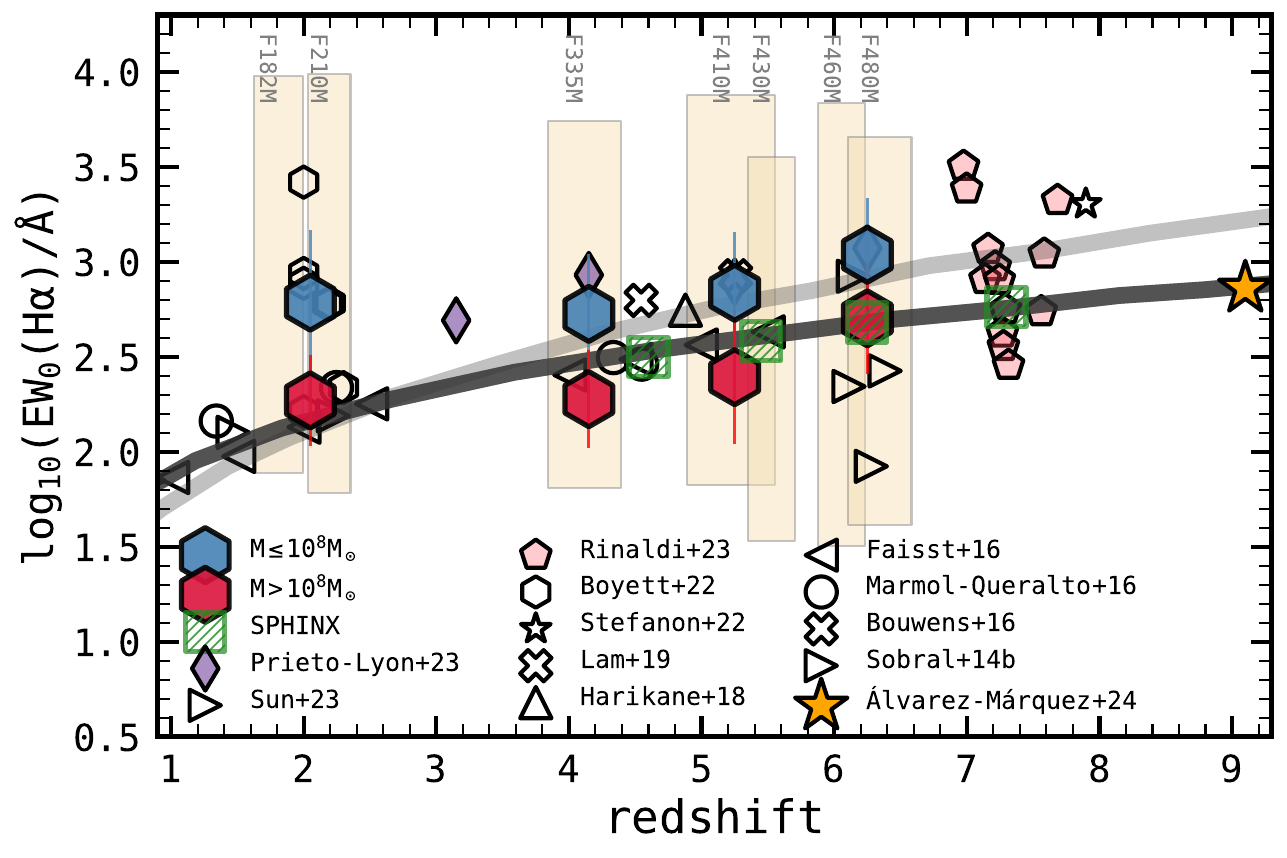}
\caption{\label{fig:ew_vs_z} Evolution of EW$_0$(H$\alpha$) against redshift. Our points are hexagons, blue for high ($\rm 10^{9.5} M_\odot > M > 10^8 M_\odot$)  and low  ($\rm  10^6 < M_\odot M < 10^8 M_\odot$) stellar masses. The light and dark grey lines represent an evolution proportional to $(1+z)^{2.1}$ and $(1+z)^{1.8}$ for $z>2.5$ and $(1+z)^{1.3}$ for $z<2.5$, respectively. Pale yellow boxes represent the redshift and EW range captured by each medium band. Literature is shown as empty markers for studies with typically higher values of stellar mass compared to our sample \citep{sobral_stellar_2014, bouwens_lyman-continuum_2016, marmol-queralto_evolution_2016, faisst_coherent_2016, harikane_silverrush_2018, lam_mean_2019, stefanon_high_2022, boyett_early_2022, sun_bursty_2023}, and filled symbols for those with a similar (but still higher) stellar mass range \citep{rinaldi_midis_2023, prieto-lyon_production_2023}. The median trend from the SPHINX cosmological simulations is included as green hatched squares.}
\end{figure*}

\begin{figure*}[t]
\centering
\includegraphics[width=0.8\textwidth]{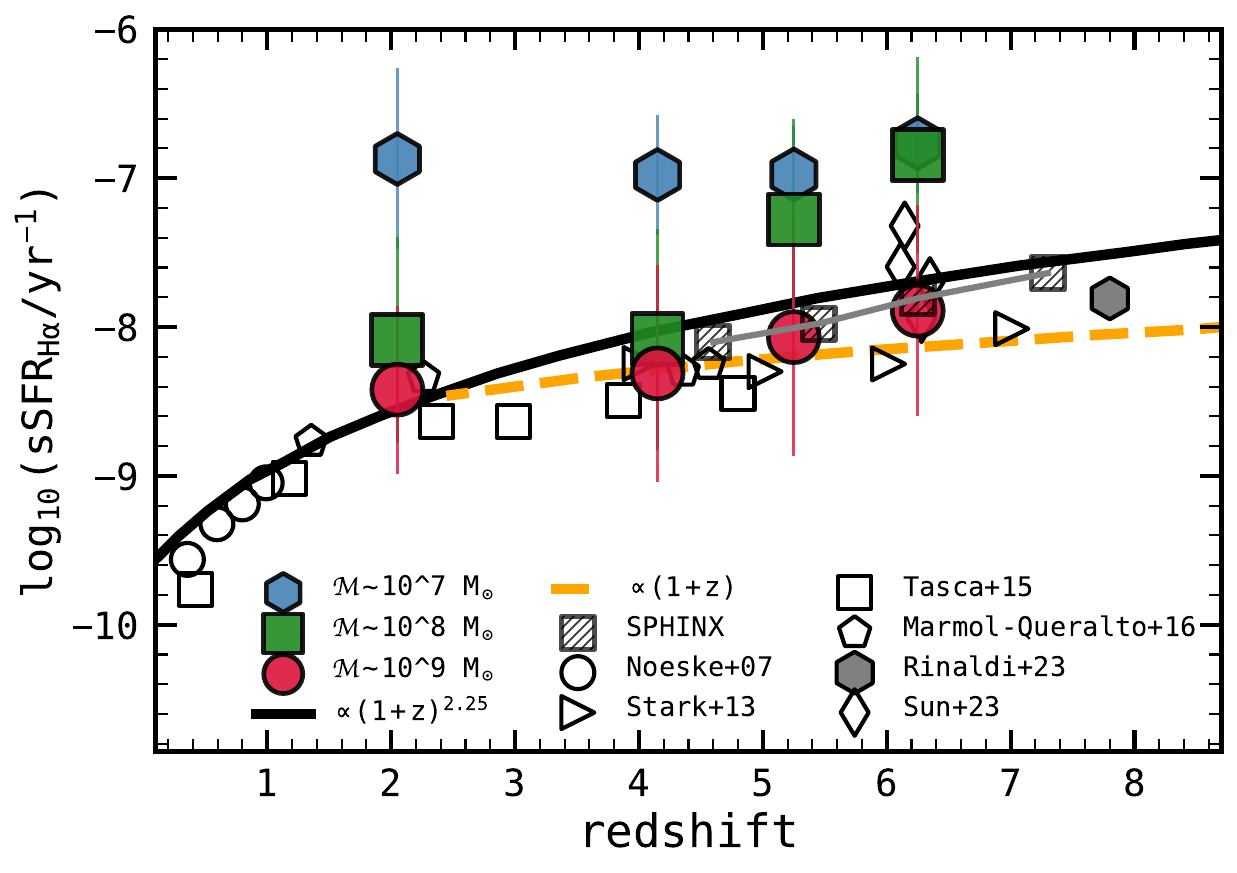}
\caption{\label{fig:ssfr_vs_z} Evolution of sSFR(H$\alpha$) as a function of redshift.  We show our data points as filled markers and color-code them by their stellar mass bin. Following the approach of \citet{marmol-queralto_evolution_2016}, we include the curves $\rm sSFR \sim (1+z)^{2.25}$ and $\rm sSFR \sim (1+z)$ as black and dashed orange lines, respectively. We overplot with black markers the literature results corresponding to sSFR from H$\alpha$ \citep{noeske_star_2007, stark_keck_2013, tasca_evolving_2015, marmol-queralto_evolution_2016, rinaldi_midis_2023, sun_first_2023}. Finally, the median values of sSFR(H$\alpha$) from SPHINX are shown as hatched squares joined by a line.}
\end{figure*}

Selection effects do not (entirely) explain the 1 dex increase in median EW$_0$ between $10^7 \, M_\odot$ and $10^9 \, M_\odot$ observed in our data. The median error for galaxies with stellar masses around $10^7 \, M_\odot$ is $0.15$ mag. The error approaches $0.05$ mag for galaxies more massive than $10^8 \, M_\odot$. 

In the right panel of Fig. \ref{fig:ew_vs_m} we show the predictions of FIRSTLIGHT \citep{ceverino_introducing_2017,ceverino_firstlight_2018,ceverino_firstlight_2019} and Sphinx galaxies \citep{rosdahl_sphinx_2018, katz_sphinx_2023} as red and blue contours, respectively. The correlation between EW$_0$(H$\alpha$) and stellar mass holds from theoretical predictions.

As can be seen in the left panel of Fig. \ref{fig:ew_vs_m}, our results at $z \sim 2$ are in good agreement with those by \citet{atek_star_2022} (shown as a grey, hatched area), although they do not completely match our stellar masses. The same happens to \citet{faisst_recent_2019}. In the case of \citet{rinaldi_midis_2024}, all their galaxies lie above our median trend, as they were selected as strong HAEs using MIRI broad bands, thus this can be explained by selection effects.

\subsection{\label{subsec:ew_versus_time} The evolution of $\rm EW_0(H\alpha)$ and $\rm sSFR(H\alpha)$ with cosmic time}

By studying the temporal evolution of EW$_0$(H$\alpha$) and $\rm sSFR(H\alpha)$ we can better understand how the properties of star-forming galaxies change across cosmic time. In particular, the evolution of EW$_0$(H$\alpha$) is closely related to the evolution of the $\rm sSFR(H\alpha)$, and both trace changes in the typical SFHs at different epochs. We investigate how EW$_0$(H$\alpha$) and $\rm sSFR(H\alpha)$ change against redshift, as we measure these quantities directly from our data.

Some previous works \citep[e.g., ][]{fumagalli_h_2012} provided functional forms for the evolution of EW$_0$(H$\alpha$) with cosmic time. These empirical relations are assumed to follow EW$_0$(H$\alpha$)$ = \epsilon (1+z)^\gamma$, motivated by simple analytical derivations, as detailed in \citealp{dekel_toy_2013}. However, the normalization factor ($\epsilon$) depends on the stellar mass range considered in each case \citep{fumagalli_h_2012, stefanon_high_2022}. 

Several studies \citep{marmol-queralto_evolution_2016,faisst_coherent_2016,stefanon_high_2022} suggested using a double power-law to provide a more robust agreement between the high and low-redshift regimes. Although they use different parametrizations,  all of them argue that the $\rm EW_0$ vs. $z$ relation is significantly steeper at $z<2.5$ compared to the one found at higher redshifts. We adopt the parametrization suggested by \citet{faisst_coherent_2016} (using pre-\textit{JWST} observations), namely: EW$_0$(H$\alpha$)$ \propto (1+z)^{1.8}$ for $z<2.5$ and EW$_0$(H$\alpha$)$ \propto (1+z)^{1.3}$ for $z>2.5$.

Later observational studies using a combination of pre-\textit{JWST} and \textit{JWST} data \citep[e.g.,][]{rinaldi_midis_2023} suggest a steeper evolution of $\rm EW_0(H\alpha)$ with $\gamma = 2.1$.

Fig. \ref{fig:ew_vs_z} shows our results for EW$_0$(H$\alpha$) versus redshift, together with a selection of works from the literature. These works span a range of stellar masses and methodologies, from spectroscopic studies \citep{faisst_coherent_2016, sun_bursty_2023, alvarez-marquez_spatially_2024} to purely photometric ones \citep{sobral_stellar_2014,marmol-queralto_evolution_2016,rinaldi_midis_2023}.

By dividing our sample around $\rm 10^8 M_\odot$ we can better match the median stellar mass of most of the literature, which is typically in the range of $\rm 10^8 M_\odot$ to $\rm 10^{10.5} M_\odot$ (e.g., \citealp{prieto-lyon_production_2023, rinaldi_midis_2023}).

We find a good agreement with the empirical relation and literature for the median trend of galaxies with $\rm M > 10^8 M_\odot$ (red hexagons). All our points are within $1 \sigma$ of the curve and the literature data points.
Galaxies with $\rm M < 10^8 M_\odot$ show an enhanced EW$_0$(H$\alpha$) for all redshift bins. This is expected, as EW$_0$(H$\alpha$) anti-correlates with stellar mass.

The points from Sphinx lie close to the empirical relation by \citet{faisst_coherent_2016} and are also compatible with our findings. Their median stellar mass is around $\rm 10^8 M_\odot$, which is compatible with the location of their median values relative to our points. However, our high-mass galaxies suffer from reduced statistics, due to the small area of JEMS compared to the full JADES area available for the rest of the bands (see Table \ref{tab:depth}).

With respect to the parametrization of the evolution of EW$_0$(H$\alpha$) versus redshift, our results are in agreement (within the error bars) with both $\gamma = 1.8$ and the steeper $\gamma = 2.1$. Moving towards even higher redshifts \citep{rinaldi_midis_2023, stefanon_high_2022, alvarez-marquez_spatially_2024}, the conclusion is also not clear-cut, and underlying systematics should be assessed.

Moving forward to the evolution of sSFR(H$\alpha$) with redshift, we find a similar picture to the one of EW$_0$(H$\alpha$), as shown in Fig. \ref{fig:ssfr_vs_z}. Parametrizations usually take the form \citep{fumagalli_h_2012} \begin{equation} \rm sSFR \sim \frac{\Dot{M}}{M} \propto M^\beta \times (1+z)^\mu,\end{equation} following the same form as the baryonic accretion rate into dark matter haloes \citep{dekel_toy_2013}. Here, $\beta$ characterizes the dependence on stellar mass and $\mu$ on redshift.

According to Eq. \ref{eq:ew_0_marmol}, our photometrically-derived $\rm EW_0(H\alpha)$ depends on stellar mass, as in general, the optical continuum becomes brighter for more massive galaxies \citep{leitherer_starburst99_1999}. For a fixed SFR (or H$\alpha$ luminosity), galaxies with higher stellar mass will have a proportionally smaller rest-frame EW compared to less massive ones. This is why the EW$_0$(H$\alpha$) is closely related to the sSFR.

By analyzing Figure \ref{fig:ssfr_vs_z}, we see that galaxies with stellar masses around $\rm \mathcal{M} \sim 10^9 \, M_\odot$ (red circles) are in good agreement with the rest of the studies (empty markers) and best-fit trends (black and orange lines). Nevertheless, the large standard deviations on the median sSFR (which in turn reflects the intrinsic scatter of values of the sSFR), prevent us from preferring $\rm sSFR \sim (1+z)^{2.25}$ or $\rm sSFR \sim (1+z)$, although our sample suggest a steeper evolution more in line with $\rm sSFR \sim (1+z)^{2.25}$, as also found by \cite{rinaldi_midis_2023} for a sample of $z\sim 7-8$ HAEs selected using MIRI photometry.

Galaxies with  $\rm \mathcal{M} \sim 10^7 \, M_\odot$ have higher median values of sSFR(H$\alpha$) than the rest of our sample, literature, and simulations. This implies that low-mass galaxies in our sample commonly experience bursts of star formation, as will be discussed in Sect. \ref{sec:results_burstiness}. 

Intermediate stellar mass galaxies have similar sSFR(H$\alpha$) compared to the more massive ones only at late cosmic times, which could follow the assembly of the main sequence of star formation for this mass range \citep{rinaldi_emergence_2024}. Selection effects could also partly be responsible for this finding, but we note that the limiting stellar mass of our sample is more than $1$ dex lower than $\rm 10^8 \, M_\odot$.

%%%%%%%%%%%%%%%%%%%%%%%%%%%%%%%%%%%%%%%%%%%%%%%%%%%%%%%%%%%%%%%%%%%%%%
%%%%%%%%%%%%%%burstiness of HAES%%%%%%%%%%%%%%%%%
%%%%%%%%%%%%%%%%%%%%%%%%%%%%%%%%%%%%%%%%%%%%%%%%%%%%%%%%%%%%%%%%%%%%%%
\section{\label{sec:results_burstiness} Evidence for bursty star formation for low-mass H$\alpha$ emitters} 

A different mode of star formation  --the so-called starbursts (SB)-- has recently been recognized as an important alternative mechanism for forming stars, especially for low stellar-mass galaxies and/or high redshifts \citep[e.g.,][]{caputi_star_2017, bisigello_analysis_2018,rinaldi_galaxy_2022, rinaldi_emergence_2024}. In this mode, the star formation rate is temporarily enhanced  (within timescales typically smaller than $100$ Myr), which in turn corresponds to a low stellar-mass doubling time. Galaxy formation models can struggle to predict starbursts, except for those that can resolve star formation down to parsec-resolution, as they can capture the small-scale physics responsible for the starbursting process (e.g., FIRE \citealp{sparre_starbursts_2017}). A more in-depth discussion on the SB phenomenon can be found in \citet{sparre_starbursts_2017, feldmann_colours_2017, ma_simulating_2018}.

Although different definitions of starbursts can be found in the literature, in this work we adopt the criterion $\rm sSFR(H\alpha) \geq 10^{-7.6}~yr^{-1}$   \citep{caputi_star_2017, caputi_alma_2021}. SB galaxies account for $54\%$ and $83\%$ of our sample at low ($z\sim 2$) and high redshifts ($4 \lesssim z \lesssim 6.5$), respectively.

For our sample of HAEs, we find that for low stellar masses ($\rm \mathcal{M} \sim 10^6 - 10^{7.25} \, M_\odot$), there is no distinction between MS or SB galaxies, independent of EW$_0$(H$\alpha$), consistent with the findings of \citet{rinaldi_emergence_2024}. This likely indicates an increased degree of burstiness in low-mass HAEs, but could be partly influenced by incompleteness in our flux-limited sample, as the $75 \%$ completeness stellar mass for JADES is $\rm \sim 10^{7.25} M_\odot$ at the highest redshift of our study.

From $\rm 10^{7.25} \, M_\odot$ and above, a bimodality between SB and MS can be seen in the SFR-$\mathcal{M}$ plane. In particular, galaxies with higher values of EW$_0$(H$\alpha$) tend to show more departure from the MS, while low EW$_0$(H$\alpha$) ones are typically located in the MS of star formation. Note that the gap in the bimodal sSFR distribution is sensitive to the choice of star-formation histories for the SED fitting model templates (as discussed in \citealp{rinaldi_emergence_2024}), but the MS/SB sequences can always be separately identified.

\begin{figure}[t]
\centering
\includegraphics[width=0.9\columnwidth]{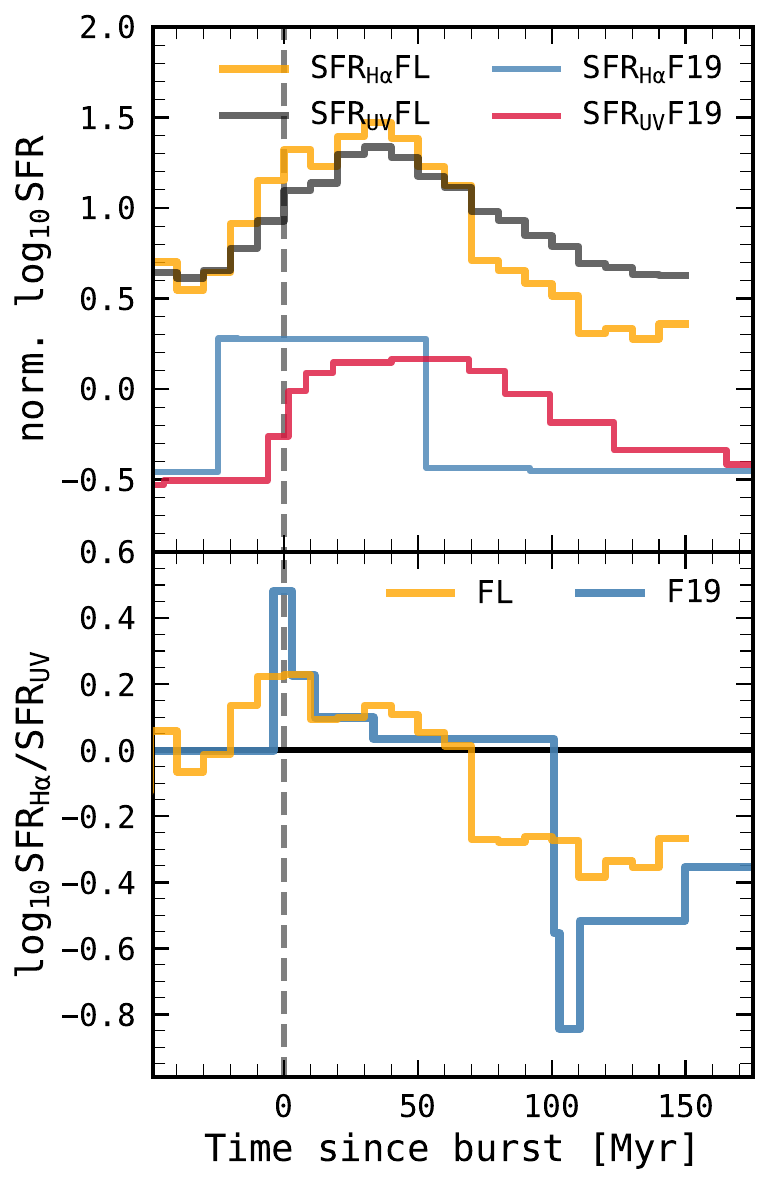}
\caption{\label{fig:ratio_simulations} \textbf{Upper panel}: H$\alpha$ and UV-derived star formation histories for an average star formation burst in FIRSTLIGHT (orange and grey lines) simulations, together with a model presented by \citet{faisst_recent_2019} (red and blue lines). The scale is arbitrary and has been shifted. \textbf{Lower panel}: ratio between H$\alpha$ and UV-derived star formation rates versus time for FIRSTLIGHT (orange) and the model from \citet{faisst_recent_2019} (blue line), showing how SFR(H$\alpha$)/SFR(UV)$>$1 during the first $\rm \sim 50$ Myrs after the burst, and then SFR(H$\alpha$)/SFR(UV)$<$1 after $\rm \sim 100$ Myrs.}
\end{figure}

In the following sections, we explore how the ratio SFR(H$\alpha$)/SFR(UV) can trace recent bursts of star formation. We will also explore the possible relation between these two tracers of burstiness: sSFR(H$\alpha$) and SFR(H$\alpha$)/SFR(UV).

\begin{figure*}[ht]
\centering
\includegraphics[width=0.9\textwidth]{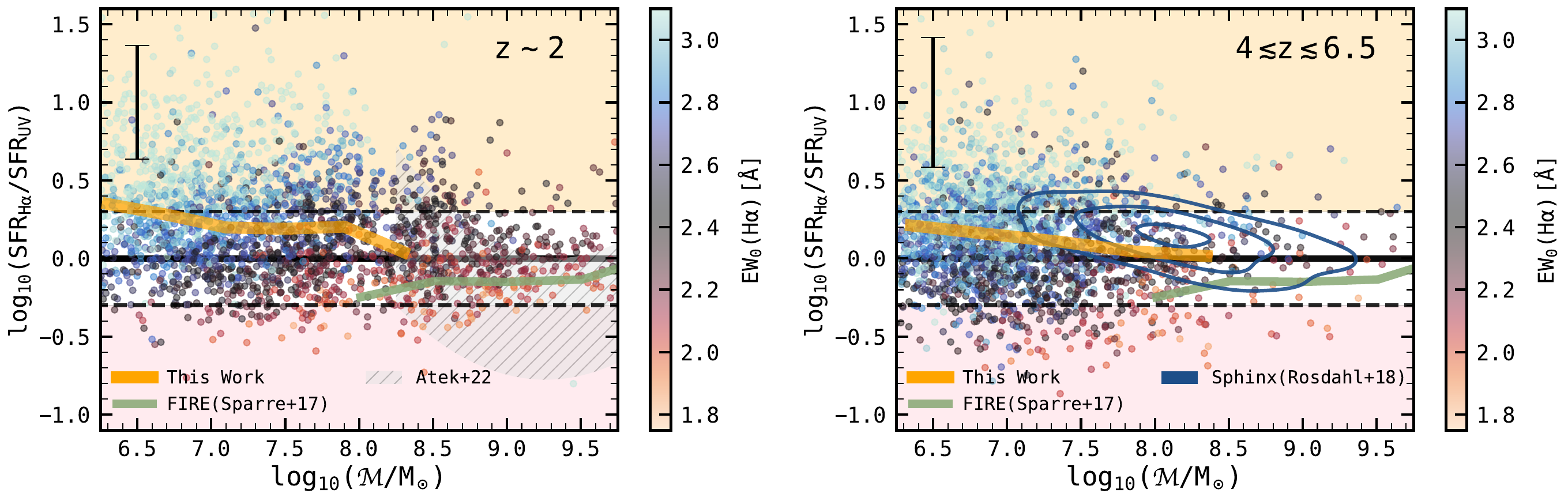}
\caption{\label{fig:ratio_ha_uv} SFR(H$\alpha$)/SFR(UV) ratio versus stellar mass for both redshift sub-samples of HAEs. We color-coded our data points by EW$\rm_0$(H$\alpha$) (high EW is colored as blue). The median trend is shown as an orange line. The results by \citet{atek_star_2022} are included in the low-redshift plot as a hatched grey contour. The median trend of FIRE \citep{sparre_starbursts_2017} is shown as a green line, while we include blue contours for SPHINX simulation \citet{rosdahl_sphinx_2018, katz_sphinx_2023}. }
\end{figure*}

\subsection{The evolution of SFR(H$\alpha$)/SFR(UV) following a burst of SF \label{sfr_ha_uv_evol}}

As shown in Fig. \ref{fig:ew_age}, the H$\alpha$ emission traces the very recent SFH. For an instantaneous burst, stellar tracks show that the EW$\rm_0$(H$\alpha$) quickly declines $2$ dex in the first $10$ Myr after the burst. This is because H$\alpha$ is linked to strong photoionization from massive and short-lived O and B stars. Thus, H$\alpha$ traces the past $\sim 10$ Myr of the SFH. Far-UV luminosity around $1500 \, \rm \AA$ is another proxy for SF, and traces timescales of $\sim 100$ Myr as described by \citet{madau_cosmic_2014}.

For this reason, the ratio between UV and H$\alpha$-derived SFR can be used to trace galaxies that have experienced a recent SF burst (e.g. \citealp{lee_comparison_2009, weisz_modeling_2012, guo_bursty_2016, atek_star_2022}). By modeling galaxies with different SFHs, it is possible to trace the behavior of SFR(H$\alpha$)/SFR(UV) ratio after a burst \citep{weisz_modeling_2012, faisst_recent_2019, emami_closer_2019}. Following the discussion by \citet{atek_star_2022}, the models suggest that:
\begin{enumerate}
    \item During the first million years and until $\sim 50$ Myr after the burst, H$\alpha$ is enhanced by the onset of star formation, while the UV luminosity is still rising. The ratio SFR(H$\alpha$)/SFR(UV) is enhanced well above unity.
    \item By $\sim 50$ Myr, O and B stars have gone out of the main sequence and died, so the H$\alpha$ emission decreases rapidly. However, the stars that contribute most to the UV emission live longer. SFR(H$\alpha$)/SFR(UV) is below unity.
    \item By $\sim 300$ Myr after the burst, these stars die as well, and the ratio SFR(H$\alpha$)/SFR(UV) returns to unity.
\end{enumerate}

Thus, galaxies whose SFR(H$\alpha$)/SFR(UV) ratio departs from unity (typically more than $\rm \pm 0.3$ dex, a threshold found in previous analysis and compatible to the uncertainty of our SFR(H$\alpha$)/SFR(UV) determination), can be associated with having a burst in their SFH. We can infer a recent burst in the case of an above-unity SFR(H$\alpha$)/SFR(UV), and a burst in the last $200$ Myr in the case of a below-unity ratio (see \citealp{faisst_recent_2019} for an in-detail discussion). The actual duration of the SF burst and the interval between bursts can affect the overall distribution of SFR(H$\alpha$)/SFR(UV) ratios, but the broad division in SFR(H$\alpha$)/SFR(UV) above or below unity still holds.

We further analyze the effect of burstiness in the SFR(H$\alpha$)/SFR(UV) ratio by using SFHs from both models and simulations in Fig. \ref{fig:ratio_simulations}. We explore a burst with a duration of $\rm \sim 100$ Myrs for both a galaxy from FIRSTLIGHT and a simple model proposed by \citet{faisst_recent_2019}.

We then extract the SFRs from the H$\alpha$ and the UV, which can be seen in the upper panel of Fig. \ref{fig:ratio_simulations}. The scale in the upper panel has been normalized and shifted to facilitate an easier visualization. Here, it is clear that for both the model and simulations, the SFR derived for the UV is lagging and smoother compared to the one from the H$\alpha$, as the latter is sensitive to variations on shorter timescales.

The lower panel of Fig. \ref{fig:ratio_simulations} shows how the SFR(H$\alpha$)/SFR(UV) ratio is larger than unity for the first $\rm 50$ Myrs after the burst. Then, it becomes smaller than unity to return to identity after some hundred million years of constant SFH.

\begin{figure*}[ht]
\centering
\includegraphics[width=1\textwidth]{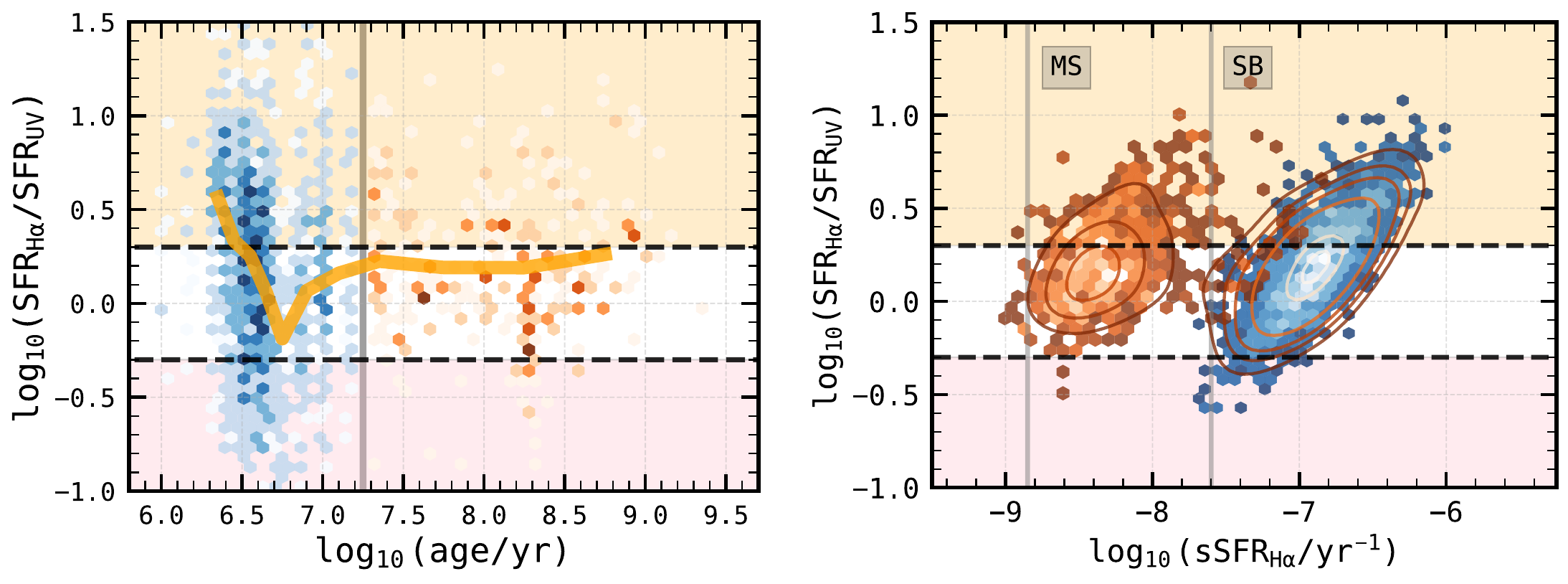}
\caption{\label{fig:ratio_panel} \textbf{Left:} SFR(H$\alpha$)/SFR(UV) versus best-fit age. Hexagons represent the kernel density estimate of the HAE sample, separated by $<10^{7.25}$ yr (young, blue) and $10^{7.25}$ yr (old, orange). The orange line represents the median trend. The orange and pink areas are defined as in Fig. \ref{fig:ratio_ha_uv}. \textbf{Right:} SFR(H$\alpha$)/SFR(UV) versus sSFR(H$\alpha$). Hexagons represent the kernel density estimate of our HAE sample, following the same color coding as in the left panel.}
\end{figure*}

\subsection{Inferring burstiness from SFR(H$\alpha$)/SFR(UV)}

In Fig. \ref{fig:ratio_ha_uv} we show the (dust corrected) SFR(H$\alpha$)/SFR(UV) ratio as a function of stellar mass. We color-code our galaxies by their EW$\rm_0$(H$\alpha$). From the behavior of the median SFR(H$\alpha$)/SFR(UV) (orange) trend in Fig. \ref{fig:ratio_ha_uv}, we determine that low-mass galaxies ($\rm \mathcal{M} < 10^8 \, M_\odot$) typically show enhanced SFR(H$\alpha$)/SFR(UV) ratios when compared to their massive counterparts. In general, galaxies with elevated values of EW$\rm_0$(H$\alpha$) tend to have the highest values of SFR(H$\alpha$)/SFR(UV) and be located within the area shaded in orange.

The running median was found to take positive values, well within the scatter of the MS (taken as $\rm \pm 0.3$ dex, and delimited by discontinuous horizontal lines in Fig. \ref{fig:ratio_ha_uv}), except for the lowest stellar masses in our sample, where it lays above it.

The uncertainties associated with our measurements of SFR(H$\alpha$)/SFR(UV) are large, especially for the lower mass galaxies. We show the median uncertainty on SFR(H$\alpha$)/SFR(UV) for all galaxies with stellar masses smaller than $\rm 10^{8} \, M_\odot$ (as shown in each of the panels of Fig. \ref{fig:ratio_ha_uv}). From here, it can be seen that the presence of galaxies with both high EW$\rm_0$(H$\alpha$) and elevated SFR(H$\alpha$)/SFR(UV) ratios only at low stellar masses can not be fully explained with larger uncertainties. 

Interestingly, we capture a non-negligible number of galaxies with SFR(H$\alpha$)/SFR(UV) ratio smaller than unity, usually for the ones with lower values of EW$\rm_0$(H$\alpha$). These values are compatible with a SF burst that happened between $\rm \sim 50-300~Myr$ in the past (Sect. \ref{sfr_ha_uv_evol}). The number of these systems is relatively small compared to the ones with SFR(H$\alpha$)/SFR(UV) unity or higher. This effect could be a consequence of our selection method (although our minimum EW$\rm_0$(H$\alpha$) is around $50 \, \rm \AA$). We emphasize that the number of galaxies with  SFR(H$\alpha$)/SFR(UV) below unity decreases only below $\rm 10^{7} \, M_\odot$, which can give an idea about the completeness of our dataset.

All in all, considering all galaxies with signatures of burstiness in their SFHs (in the orange or red shaded areas of Fig. \ref{fig:ratio_ha_uv}), the fraction of low-mass galaxies ($\rm \mathcal{M} < 10^{8} \, M_\odot$) with signs of bursty star formation is $\rm \sim 50 \%$. It quickly drops to $\rm \sim 25 \%$ for $\rm \mathcal{M} > 10^{8} \, M_\odot$. This is consistent with the values from the $\rm sSFR(H\alpha)$, showing SB fractions of $80\%$ and $17\%$, respectively.

The emerging picture from this results is in line with observational studies that asses the recent changes in the SFH of galaxies using similar methods as in our study \citep[e.g.,][]{asada_bursty_2024,endsley_burstiness_2024}. Namely, low-mass galaxies ($\rm \mathcal{M}<10^{8}~M_\odot$) commonly show enhanced SFR(H$\alpha$)/SFR(UV) ratios, relative to their massive counterparts. By leveraging the extremely deep JADES photometry, we are able to confirm this for HAEs with stellar masses previously inaccessible ($\rm \mathcal{M}\sim10^{7}~M_\odot$) at $z\sim6$.

The relation between two indicators of burstiness (SFR(H$\alpha$)/SFR(UV) and sSFR(H$\alpha$) is shown in the right panel of Fig. \ref{fig:ratio_panel}. As from our previous findings, young galaxies (that conform to the SB cloud, as defined from their sSFR(H$\alpha$), are the bulk of our galaxy sample and shows a wider distribution of SFR(H$\alpha$)/SFR(UV) values, which reflects the intrinsic effects of burstiness. Not all galaxies that fall in the SB cloud would be classified as bursty from their elevated SFR(H$\alpha$)/SFR(UV) ratio, as this value is strongly dependent on time since the onset of star formation, and the large associated uncertainties in the determination of SFR(H$\alpha$)/SFR(UV) make the ratio only sensitive to the most recent/strongly SB systems. Accordingly, we find that around half of all galaxies considered bursty from their sSFR(H$\alpha$) are also selected from their SFR(H$\alpha$)/SFR(UV) ratio.

In the left panel of Fig. \ref{fig:ratio_panel} we illustrate the behavior of the SFR(H$\alpha$)/SFR(UV) ratio against best-fit age. The evolution of the ratio with age is closely related to the one discussed in Fig. \ref{fig:ratio_simulations}. On average, our HAE sample shows a strong anti-correlation between SFR(H$\alpha$)/SFR(UV) and age for the first $\sim 10$ Myrs (as expected from the H$\alpha$ timescales), and the younger galaxies typically are classified as bursty. By observing the kernel density estimate, most of our sample has best-fit ages $< 10^{7.25}$ Myr. This group of HAEs shows a wide distribution of SFR(H$\alpha$)/SFR(UV) values, as a consequence of the presence of burstiness in their SFHs.

Finally, we find that ratio SFR(H$\alpha$)/SFR(UV) is strongly correlated with EW$\rm_0$(H$\alpha$). In our sample, galaxies with EW$\rm_0$(H$\alpha$) $\rm \gsim 2.75 \AA$  (which account for $\rm 51 \%$ of our total sample) show over-unity SFR(H$\alpha$)/SFR(UV) ratios, and an average $\rm SFR(H\alpha)/SFR(UV) = 0.45 \pm 0.34$. Their average $\rm log_{10}$sSFR(H$\alpha$) is $\rm -6.87 \pm 0.54$, above the threshold we use for defining SB galaxies ($\rm log_{10}$sSFR(H$\alpha$)$>-7.6$).
667

Now, we put our results in the context of previous observational and theoretical studies.  \citet{atek_star_2022} find that SFR(H$\alpha$)/SFR(UV) is above-unity for galaxies with $\rm \mathcal{M} < 10^{8.5} M_\odot$. However, the lack of galaxies below unity at low stellar masses could have originated from incompleteness, creating an artificially steep increase of the median trend shown in Figure \ref{fig:ratio_ha_uv}. They also report an increase in the scatter of SFR(H$\alpha$)/SFR(UV) toward lower stellar masses, which we can also find in our sample. Following their discussion, a stochastic sampling of the initial mass function (IMF) \citep{fumagalli_stochastic_2011} could produce deviations from unity SFR(H$\alpha$)/SFR(UV), as H$\alpha$ is especially sensitive to the number of massive stars. This effect could be particularly important for galaxies with stellar masses as low as $\rm 10^6 \, M_\odot$.

Star formation histories from simulations predict that burstiness is the dominant mode of star formation for low-mass galaxies. FIRE zoom-in simulations \citep{sparre_starbursts_2017} predict that the SF happens in bursts that are generally followed by an almost complete suppression of SF by intense stellar feedback-driven outflows after $\sim 100$ Myr. The prevalence of bursty SFHs grows with decreasing stellar mass.

By using FIRSTLIGHT zoom-in simulations, \citet{ceverino_firstlight_2018} find that, on average, galaxies of stellar masses $\rm 6\leq \mathcal{M} \leq 8$ spend $70\%$ of their time undergoing SF bursts (at $z\sim 5$). Here, SF bursts reach a sSFR of $5-15$~Gyr (galaxies from FIRSTLIGHT do not populate the SB cloud) and have an average duration of $100$~Myr. The young ages we find in our work, typically around tens of million years, are compatible with galaxies that have experienced a very recent SF burst.

However, current zoom-in simulations do not yet converge on a common picture for the starburst phase, perhaps because of different implementations of star formation and feedback, together with (in some cases) a lack of resolution to model the star formation in full detail. 

\section{Summary and Conclusions} \label{sec:conclusion}
%%%%%%%%%%%%%%%%%%%%%%%%%%%%%%%%%%%%%%%%%%%%%%%%%%%%%%
%%%%%%%%%%%%%%%%  CONCLUSIONS %%%%%%%%%%%%%%%%%%%%%%%%
%%%%%%%%%%%%%%%%%%%%%%%%%%%%%%%%%%%%%%%%%%%%%%%%%%%%%%

In this paper, we studied the properties of HAEs from $z\sim 2$ to $z\sim 6$ and validated the line-emitters photometric-excess selection technique using deep medium-band JADES photometry. We constructed a statistical sample of HAEs with stellar masses as low as $\rm \sim 10^{6.5} M_\odot$ up to $z \sim 6$, facilitated by the remarkable sensitivity of medium-band photometry from \textit{JWST}/NIRCam JADES observations. Currently, \textit{JWST} is the only facility that allows us to prove the optical rest frame for galaxies up to $z\sim 8$. We emphasize that no previous study has explored a statistical sample of HAEs at redshifts $z>4$ while reaching stellar masses as low as $\rm \sim 10^{6.5} M_\odot$, however, we highlight recent \textit{JWST}-based analysis that are starting to explore a similar parameter space \citep[e.g.][]{endsley_burstiness_2024, asada_bursty_2024}.

We selected a sample of galaxies with signatures of emission lines in their photometry, as explored in \citet{williams_jems_2023}. We then measured the rest-frame H$\alpha$ equivalent width, H$\alpha$ flux, and finally their star formation rates from the photometric excess with respect to the best-fit model. These close to $4500$ galaxies are characterized by a mean stellar mass of $\rm 10^{7.3} \, M_\odot$ and a mean equivalent width of $\sim 900 \, \AA$, with a wide distribution of values.

In Section \ref{sec:methods} we confirmed that our method can recover H$\alpha$ line fluxes by comparing them against NIRSpec/MSA line flux measurements from JADES \citep{deugenio_jades_2024}. We found that both values typically agree within one standard deviation. This confirms that no clear systematic effect biases our emission line fluxes measurements, at least for the subsample of galaxies with spectroscopic data.

In the following sections, we analyzed the properties of the H$\alpha$ emission and the host galaxies. In Section \ref{subsec:ew_versus_mass}, we find an anti-correlation between stellar mass and EW$\rm_0$(H$\alpha$), which was already reported for higher stellar masses in previous studies. The anti-correlation, however, tentatively appears to flatten for low stellar masses, saturating around $\rm 10^{6.5-7} \, M_\odot$.

When studying the evolution of EW$\rm_0$(H$\alpha$) with cosmic time (Section \ref{subsec:ew_versus_time}), we find that the median trend of our sample is compatible with an evolution of the shape EW$_0$(H$\alpha$)$ = \epsilon (1+z)^\gamma$, as also found in literature. Our results are compatible with $\rm \gamma \sim 2.1$, but do not rule out a shallower evolution. We emphasize the influence of the stellar mass on the median EW$\rm_0$(H$\alpha$) of a galaxy sample and on the normalization of the parameterization of its evolution with cosmic time. The bulk of our sample is conformed by galaxies with low stellar mass $\rm \sim 10^{7.3} \, M_\odot$, and shows higher median values for each redshift bin compared to previous studies probing higher mass (brighter) galaxies.

By scrutinizing the SFR in the last $10$ Myr (as derived from H$\alpha$) we conclude that all the strongest emitters (with $\rm EW_0(H\alpha) \sim 1000 \, \AA $ or greater) are located in the starburst cloud, following the definition of \cite{caputi_alma_2021}. These systems also have young best-fit ages (of a couple million years on average). The evolution of the specific star formation rate behaves differently for galaxies with different stellar masses within our sample. 

Interestingly, as shown in Fig. \ref{fig:ssfr_vs_z}, galaxies with $\rm M \sim 10^{7} \, M_\odot$ tend to have high sSFRs for all cosmic times, with no evolution whatsoever. The overall values of sSFR for more massive galaxies ($\rm M \gtrsim 10^{9} \, M_\odot$), are lower and evolve with cosmic time following the trends found in previous literature. As discussed in the main text, we have asses that incompleteness is not playing an important role in any of this stellar masses, across all redshift ranges shown in the figure.

In Section \ref{sec:results_burstiness} we explore the SFHs of HAEs by comparing star formation rates derived from UV and H$\alpha$, tracers of star formation at different timescales \citep{kennicutt_star_1998}. Significant departures from unity in the factor SFR(H$\alpha$)/SFR(UV) are most likely due to a burst of star formation \citep{atek_star_2022}. In particular, an enhanced value of SFR(H$\alpha$)/SFR(UV) can be explained by a star formation burst in the past $50$ Myr.

The median SFR(H$\alpha$)/SFR(UV) in our sample is above but compatible with unity and increases towards low stellar masses, but we find an important scatter around these values. Furthermore, we find a correlation between $\rm EW_0(H\alpha)$ and an enhanced SFR(H$\alpha$)/SFR(UV) for the galaxies in our sample, which supports the scenario of recent bursty star formation for galaxies with an above-unity SFR(H$\alpha$)/SFR(UV) during the first $\rm 50$ Myr, then below-unity SFR(H$\alpha$)/SFR(UV) during the next $\rm \sim 200$ Myr.

This behavior has been recovered in simulations \citep{ceverino_firstlight_2018} and reported in previous studies \citep{faisst_recent_2019, atek_star_2022}, again for typically more massive galaxies than the ones studied here. We extract an SF burst from FIRSTLIGHT and compare it with the predictions from a model presented by \citet{faisst_recent_2019}. Both show the expected evolution of SFR(H$\alpha$)/SFR(UV) with time after an SF burst. We set a threshold of $\rm 0.3$ dex above or below unity as a proxy for the presence of an SF burst in the SFH, based on previous studies and on the uncertainties in our determination of SFR(H$\alpha$)/SFR(UV).

At last, the scatter of SFR(H$\alpha$)/SFR(UV) is present for most of our sample, especially for galaxies with stellar masses smaller than $\rm 10^{8} \, M_\odot$, and asymmetric towards greater than unity values. Taken at face value, these results could indicate that star formation in such low-mass systems is characterized by short timescales, and is highly unstable: with periods of star forming bursts followed by fast cessation of star formation \citep{sparre_starbursts_2017, ceverino_firstlight_2018}. Uncertainties in our photometrically-recovered SFR(H$\alpha$)/SFR(UV) ratio do not fully explain this behavior.

In particular, we find the fraction of low-mass systems ($\rm \mathcal{M} < 10^{8} \, M_\odot$) with signs of SF bursts (from their SFR(H$\alpha$)/SFR(UV)) to be $\rm \sim 50 \%$. This fraction quickly drops to $\rm \sim 25 \%$ for $\rm \mathcal{M} > 10^{8} \, M_\odot$. This results are consistent with what is found by using sSFR(H$\alpha$) ($\rm \sim 80 \%$ and $\rm \sim 17 \%$, respectively). Other recent studies that employ \textit{JWST} to study burstiness using a similar methodology but probing more massive galaxies \citep[e.g.,][]{asada_bursty_2024, endsley_burstiness_2024} show qualitatively compatible results.

Various other works have studied the effect of uncertainties that photometric studies can have in estimating EW$_0$(H$\alpha$) and SFR(H$\alpha$). In particular, \citet{faisst_recent_2019} concluded that metallicity, differential dust extinction between stars and gas together with the flavor of dust attenuation law can have an effect in the values of photometrically derived EW$_0$(H$\alpha$). However, the net effect of all uncertainties was found to be $\lesssim 20 \%$ for most galaxies (except for the dustiest ones, with $\rm E(B-V) \geq 0.3$, which account for less than $\rm 9 \%$ of our total sample).

As a sanity check, we implemented several star formation histories and differential attenuation values. We compared our fiducial values (differential dust extinction $f=0.44$ and exponentially declining star formation histories) against delayed SFH and no differential dust extinction ($f=1$). In all cases, results are within the estimated errors and trends still hold for all combinations of SFHs and $f$. Remarkably, we find that the median SFR(H$\alpha$)/SFR(UV) is sensitive to the adopted differential dust extinction values, getting closer to unity for $f=1$. However, the large scatter and the shape of the found trends are still consistent, especially towards small stellar masses, which indicates that our conclusions about burstiness and temporal evolution of line emission do not arise from the above assumptions.

Future studies using \textit{JWST} spectroscopic data will mitigate uncertainties on the differential dust extinction for high-redshift galaxies, and help us understand in detail the optimal way to model these systems to minimize the effect of biases that arise from our current modeling assumptions.

All in all, \textit{JWST}/NIRCAM photometry appears to be an excellent tool to probe a range of physical properties of early galaxies, especially for intrinsically faint or low stellar mass systems that are increasingly difficult to target with spectroscopic observations, even with \textit{JWST}. Nevertheless, these are needed to minimize the uncertainties inherent to photometric studies like this one, such as the presence of differential dust extinction, or the effect of IMF and metallicity, among others.

\section{ACKNOWLEDGMENTS} \label{sec:ACKNOWLEDGMENTS}
%%%%%%%%%%%%%%%%%%%%%%%%%%%%%%%%%%%%%%%%%%%%%%%%%%%%%%
%%%%%%%%%%%%%%%%  ACKNOWLEDGMENTS %%%%%%%%%%%%%%%%%%%%
%%%%%%%%%%%%%%%%%%%%%%%%%%%%%%%%%%%%%%%%%%%%%%%%%%%%%%

RNC, KIC and VK acknowledge funding from the Dutch Research Council (NWO) through the award of the Vici Grant VI.C.212.036. EI and KIC acknowledge funding from the Netherlands Research School for Astronomy (NOVA). This work is based on observations made with the NASA/ESA/CSA James Webb Space Telescope. The data were obtained from the Mikulski Archive for Space Telescopes at the Space Telescope Science Institute, which is operated by the Association of Universities for Research in Astronomy, Inc., under NASA contract NAS 5-03127 for \textit{JWST}. These observations are associated with \textit{JWST} programs GTO \#1180, GO \#1210, GO \#1963 and GO \#1895. The authors acknowledge the team led by coPIs D. Eisenstein, N. Luetzgendorf, C. Williams and P. Oesch for developing their respective observing programs with a zero-exclusive-access period. Also based on observations made with the NASA/ESA Hubble Space Telescope obtained from the Space Telescope Science Institute, which is operated by the Association of Universities for Research in Astronomy, Inc., under NASA contract NAS 526555.

\textit{Software}: \texttt{AstroPy} \citep{collaboration_astropy_2022}, \texttt{dustmaps} \citep{green_dustmaps_2018}, \texttt{LePHARE} \citep{arnouts_lephare_2011}, \texttt{Matplotlib} \citep{hunter_matplotlib_2007}, \texttt{NumPy} \citep{harris_array_2020}, \texttt{Photutils} \citep{bradley_astropyphotutils_2022}, \textsc{Python} \citep{van_rossum_python_1995}, \texttt{SciPy} \citep{virtanen_scipy_2020}, \texttt{Source Extractor} \citep{bertin_sextractor_1996}, \texttt{TOPCAT} \citep{taylor_topcat_2017}, \texttt{WebbPSF} \citep{perrin_updated_2014}.

\textit{Facilities}: \textit{HST}, \textit{JWST}

%%%%%%%%%%%%%%%%%%%%%%%%%%%%%%%%%%%%%%%%%%%%%%%%%%%%%%
%%%%%%%%%%%%%%%%%%%%%  APENDIX %%%%%%%%%%%%%%%%%%%%%%%
%%%%%%%%%%%%%%%%%%%%%%%%%%%%%%%%%%%%%%%%%%%%%%%%%%%%%%
\newpage
\appendix

\section{Depth of the medium bands used in our analysis \label{appendix}}

\begin{figure*}[ht]
\centering
\includegraphics[width=0.8\textwidth]{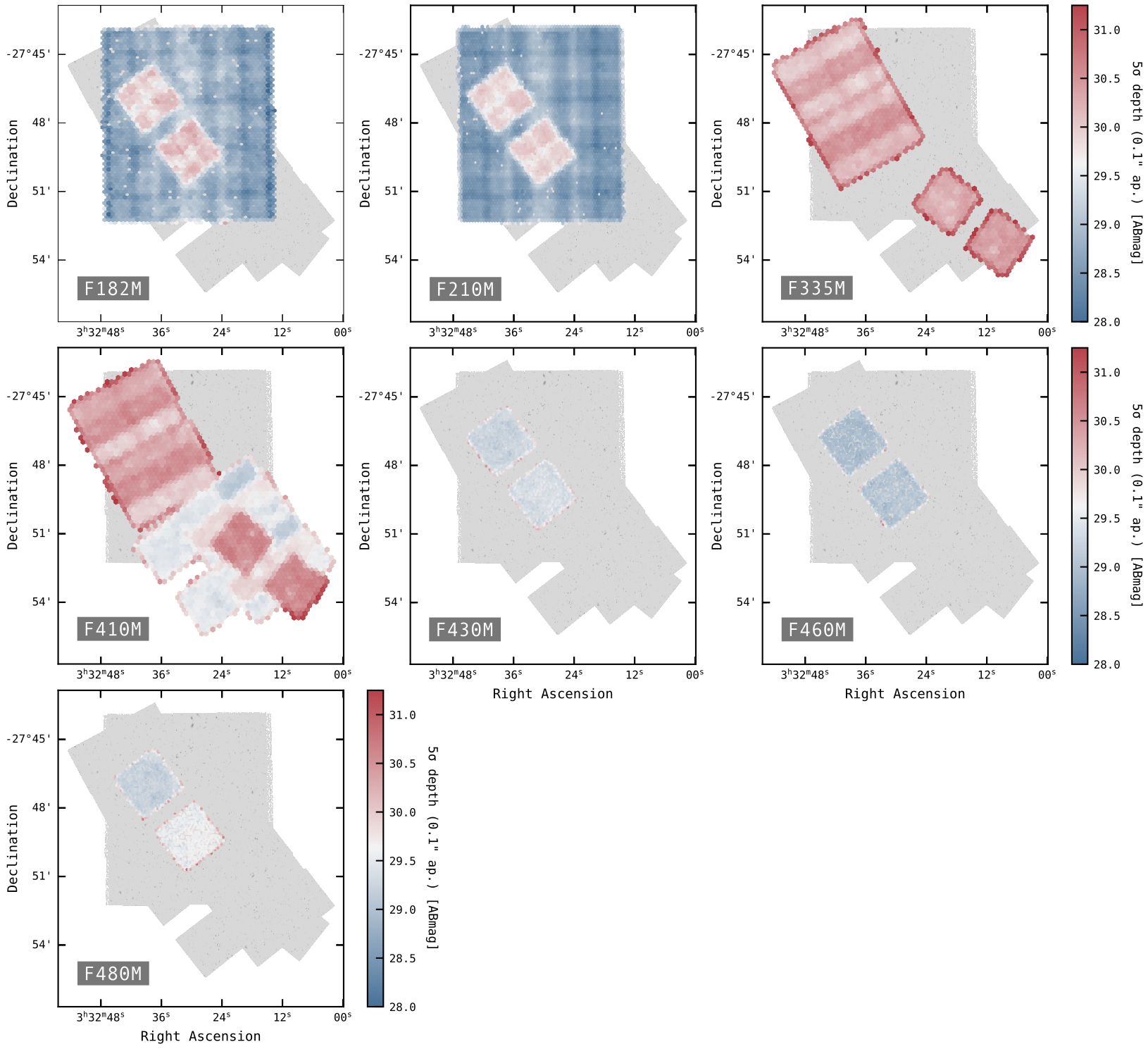}
\caption{Local depth map of all medium bands used to compute the photometric excess. The colormap represents the $5 \sigma$ image depth, measured in 0".25 (radius) apertures. Note how bands with JEMS only (e.g. F430M, F460M and F480M) have a significant different footprint from the rest of bands.}
\label{fig:local_depth}
\end{figure*}

Fig. \ref{fig:local_depth} depicts the local ($5 \sigma$) depth of all medium bands used in our study. This depths have been measured by placing a series of random $0.\arcsec 1$ apertures in blank parts of the sky within $20\arcsec \times 20 \arcsec$ of each source.

%%%%%%%%%%%%%%%%%%%%%%%%%%%%%%%%%%%%%%%%%%%%%%%%%%%%%%
%%%%%%%%%%%%%%%%%%%%  REFERENCES %%%%%%%%%%%%%%%%%%%%%
%%%%%%%%%%%%%%%%%%%%%%%%%%%%%%%%%%%%%%%%%%%%%%%%%%%%%%
\bibliography{references}{}
\bibliographystyle{aasjournal}

\end{document}